\documentclass[letterpaper,aps,prd,twocolumn,tightenlines,preprintnumbers,nofootinbib,showkeys,superscriptaddress]{revtex4-1}
\pdfoutput=1

\usepackage{XCharter}
\usepackage[T1]{fontenc}
\usepackage{mathptmx}
\usepackage{mathtools}
\usepackage{eqnarray}
\usepackage[normalem]{ulem}
\usepackage{fullpage}
\usepackage{amsfonts}
\usepackage{amsmath}
\usepackage{slashed}
\usepackage{amssymb}
\usepackage{graphicx}
\usepackage{epic}
\usepackage{eepic}
\usepackage{epsfig}
\usepackage{latexsym}
\usepackage[dvipsnames]{xcolor}
\usepackage[export]{adjustbox}
\usepackage{float}
\usepackage{multirow}
\usepackage{hyperref}
\usepackage{enumitem}
\hypersetup{colorlinks=true,citecolor=red,linkcolor=NavyBlue,urlcolor=NavyBlue}
\usepackage[caption=false]{subfig}
 
\usepackage{natbib}
\usepackage{relsize}
\usepackage[left=1.75cm,right=1.75cm,top=2.cm,bottom=2.cm]{geometry}
\usepackage{adjustbox}

\usepackage{shorthand}

\begin{document}

\title{Search for the $Z^\prime$ boson decaying to a right-handed neutrino pair in leptophobic $\mathrm{U(1)}$ models}

\author{Mathew Thomas Arun}
\email{mathewthomas@iisertvm.ac.in}
\affiliation{Indian Institute of Science Education and Research Thiruvananthapuram, Vithura, Kerala, 695 551, India} 

\author{Arpan Chatterjee}
\email{arpan.chatterjee@ut.ee}
\affiliation{Indian Institute of Science Education and Research Thiruvananthapuram, Vithura, Kerala, 695 551, India}
\affiliation{Institute of Physics, University of Tartu, W. Ostwaldi 1, EE-50411 Tartu, Estonia}

\author{Tanumoy Mandal}
\email{tanumoy@iisertvm.ac.in}
\affiliation{Indian Institute of Science Education and Research Thiruvananthapuram, Vithura, Kerala, 695 551, India}

\author{Subhadip Mitra}
\email{subhadip.mitra@iiit.ac.in}
\affiliation{Center for Computational Natural Sciences and Bioinformatics, International Institute of Information Technology, Hyderabad 500 032, India}

\author{Ananya Mukherjee}
\email{ananyatezpur@gmail.com}
\affiliation{Theory Division, Saha Institute of Nuclear Physics, 1/AF Bidhannagar, Kolkata
700 064, India} 

\author{Krishna Nivedita}
\email{knivedita@science.ru.nl}
\affiliation{Indian Institute of Science Education and Research Thiruvananthapuram, Vithura, Kerala, 695 551, India}
\affiliation{Department of astrophysics/IMAPP, Radboud University, P.O Box 9010, 6500 GL Nijmegen, The Netherlands}
\date{\today}
             
\begin{abstract} 
\noindent 
The $U(1)$ extensions of the Standard Model contain a heavy neutral gauge boson $Z^\prime$. If leptophobic, the boson can evade the stringent bounds from the dilepton resonance searches. We consider two theoretically well-motivated examples of leptophobic $U(1)$ extensions in which the $Z'$ decays to right-handed neutrinos (RHNs) with substantial branchings. The coexistence of a leptophobic $Z^\prime$ and the RHNs opens up a new possibility of searching for these particles simultaneously through the production of a $Z^\prime$ at the LHC and its decay to a RHN pair. For this decay to occur, the RHNs need to be lighter than the $Z^\prime$. Hence, we study this process in an inverse seesaw setup where the RHNs can be in the TeV range. However, in this case, they  have a pseudo-Dirac  nature, i.e., a RHN pair would produce only opposite-sign lepton pairs, as opposed to the Majorana-type neutrinos, which can produce both same- and opposite-sign lepton pairs. Hence, the final state we study has a same-flavour opposite-sign lepton pair plus hadronically-decaying boosted $W$ bosons. Our analysis shows that the high luminosity LHC can discover a TeV-scale leptophobic $Z^\prime$ decaying via a RHN pair in a wide range of available parameters. Interestingly, large parameter regions beyond the reach of future dijet-resonance searches can be probed exclusively through our channel.
\end{abstract}

\maketitle 
\section{Introduction}
\label{section:introduction}
\noindent
One of the simple ways to extend the Standard Model (SM) nontrivially is to append a local $U(1)$ group to its gauge structure. The $U(1)$ extensions are well-motivated in both top-down theories and bottom-up models, and the associated literature is vast. A remnant $U(1)$ symmetry that is broken at about a few TeV leaving a massive vector boson $Z^\prime$ can come from the grand unified theories (GUTs)~\cite{Langacker:1980js,London:1986dk,Hewett:1988xc}. Reviews on the phenomenological aspects of $Z^\prime$ are found in~\cite{Leike:1998wr,Langacker:2008yv} and references therein. 

Nonobservation of a $Z^\prime$ resonance in the conventional search channels at the LHC impels us to look for new channels that are theoretically well-motivated. In this paper, we consider a bottom-up model of a TeV-scale $Z^\prime$ in a $U(1)$ extension to study an unexplored search channel of $Z^\prime$ through its decay to a pair of right-handed neutrinos (RHNs, $N_R$'s). Adding a $U(1)$ gauge group to the SM can lead to gauge anomalies breaking the gauge invariance and renormalisability of the theory. These anomalies can be cancelled by introducing new chiral fermions such as the RHNs~\cite{Ekstedt:2016wyi} or the Green-Schwarz (GS) mechanism~\cite{Leontaris:1999wf,Ekstedt:2017tbo}. Therefore, RHNs are present in most anomaly-free $U(1)$ models. 

If the branching ratios (BRs) of $Z'$ to RHN pairs are large, they can provide a complimentary search channel of $Z^\prime$ leading to a discovery at the high luminosity LHC (HL-LHC). The channel is also important from the RHN-search point of view. Since the RHNs are singlets under the SM gauge group, they can be produced from the SM fields only through their overlaps with the SM neutrinos, which are small. However, if a new particle decays to the RHNs, the corresponding cross section can be large enough to observe the process at the LHC.


So far, the LHC has looked for a TeV-scale $Z^\prime$ in the dijet~\cite{ATLAS:2019fgd,CMS:2019gwf}, dilepton~\cite{ATLAS:2019erb,CMS:2021ctt}, diphoton~\cite{ATLAS:2016gzy,CMS:2016kgr}, diboson~\cite{ATLAS:2020fry,CMS:2021klu}, and  $t\bar{t}$~\cite{ATLAS:2012dgv,CMS:2015fhb} channels (see Ref.~\cite{ATLAS:2018tvr} for the prospects of $Z^\prime$ searches in the dilepton channel at the HL-LHC); but not in the RHN channel. Among the searched ones, the dilepton resonance searches put the best limits on the $Z^\prime$ mass. For example, a sequential $Z^\prime$ (whose couplings to the SM fermions are the same as the $Z$ boson) is excluded up to $\sim 5$~TeV~\cite{ATLAS:2019erb}. However, if the  $Z^\prime$ is leptophobic, it would not couple to the SM leptons, and hence, it can still exist at a relatively lower mass range since the bounds from the other channels (like the dijet) are not so severe. 

For our purpose, then,  we consider a leptophobic $Z^\prime$ that largely decays to RHNs. The RHN channel has been discussed in some phenomenological contexts earlier~\cite{Ferrari:2002ac,Das:2017flq,Das:2017deo,Cox:2017eme,Das:2022rbl}, but not in a leptophobic setup. Thus, we focus on the part of the $U(1)$ parameter space that is unexplored both experimentally and phenomenologically so far.
A low-scale leptophobic $U(1)$ can be realised in the GUT models~\cite{Babu:1996vt,Lopez:1996ta,Rizzo:1998ut,Leroux:2001fx}. It is also possible that the RHNs only cancel the mixed $U(1)$ gauge-gravity anomalies in the triangle diagrams, and some other mechanisms like the GS mechanism cancel the rests. We discuss these possibilities in the next section.

\begin{figure}
\includegraphics[width=0.9\columnwidth]{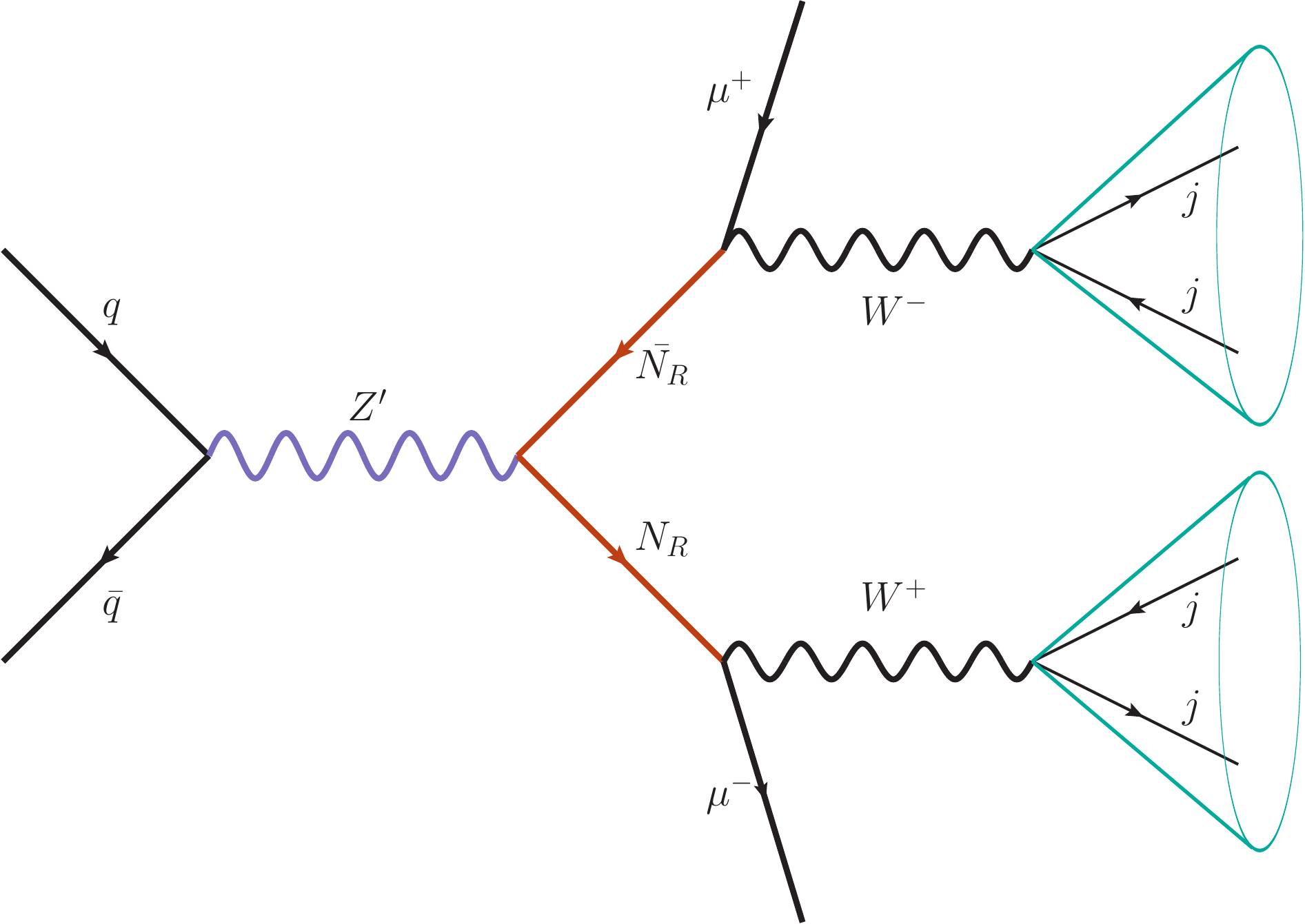}
\caption{A representative Feynman diagram of the signal process leading to an opposite-sign muon pair and $W$-like fatjets in the final state. In our analysis, we consider only muons for their high detector sensitivity.}
\label{fig:FD}
\end{figure} 

The RHNs can be of Majorana or Dirac type depending on the seesaw mechanism in play. In the standard type-I seesaw mechanism~\cite{Minkowski:1977sc,Mohapatra:1979ia}, one introduces very heavy ($\sim 10^{14}$~GeV for order one Yukawa couplings) Majorana-type RHNs to explain the observed tiny masses of the light neutrinos. In contrast, with the inverse-seesaw mechanism (ISM)~\cite{Mohapatra:1986aw,Mohapatra:1986bd}, they are pseudo-Dirac type but lighter---about the TeV-scale, i.e., within the reach of the LHC. Since we are interested in a $Z^\prime$ decaying to a RHN pair within the reach of the LHC, we consider the ISM. Discussions on various phenomenological aspects of $U(1)$ extensions and ISM can be found in Refs.~\cite{Das:2017kkm,Jana:2019mez,Das:2019pua,Choudhury:2020cpm,Bandyopadhyay:2020djh,Deka:2021koh,Das:2021nqj}. Heavy neutrino searches at future lepton colliders can be found in~\cite{Banerjee:2015gca,Chakraborty:2018khw,Barducci:2022hll}.

In general, the $pp\to Z^\prime\to N_RN_R$ process can lead to various final states through different decays of the RHN pair.
We focus on those producing a same-flavour  lepton pair and $W$-like fatjets (see Fig.~\ref{fig:FD}).
When the RHNs are Majorana-type fermions, they can produce both same-sign (SS) and opposite-sign dilepton (OSDL) final states~\cite{Han:2006ip}. The SSDL channel is a clean probe of Majorana-type RHNs because of the low SM background. However, since the Dirac-type RHNs would produce only OSDL final states,  the SSDL probe would not work in presence of the ISM. In other words, the presence of OSDL events but no SSDL event can hint towards Dirac-type RHNs~\cite{Chen:2011hc,Das:2017hmg} and hence, the ISM. Isolating the OSDL channel from its large SM background, however, is a lot more challenging compared to the SSDL channel. Here, we shall investigate the prospects of the OSDL channel as the signature of the ISM within a leptophobic $U(1)$ set-up.

The paper is organised as follows: In Sec.~\ref{sec:LPU1}, we briefly discuss the possible leptophobic $U(1)$ constructions; in Sec.~\ref{sec:zpbound}, we introduce the phenomenological $Z'$ model and discuss the latest LHC bounds on the parameter space; we study the signatures arising from the pair production of RHNs in Sec.~\ref{sec:sigpro}; in Section~\ref{sec:sigback}, we discuss the signal and relevant SM background processes; finally, we conclude in Sec.~\ref{sec:conclu}.

\section{Leptophobic $\mathrm{U}(1)$ extensions}
\label{sec:LPU1}

\noindent
As discussed in the Introduction, we are interested in a TeV-range leptophobic $Z'$ that resonantly decays to a RHN pair. A $Z'$ is naturally present in gauged $U(1)$ extensions of the SM as the mediator of the new force. In this section, we theoretically motivate our desired scenario  with two examples. 

\subsection{A model with the GS mechanism}\label{subsec:gs}\noindent
We extend the SM gauge sector by an additional Abelian gauge group, $U(1)_z$ and introduce operators to cancel the new gauge anomalies through the GS mechanism. 
While the  GS mechanism normally does not require any additional chiral fermion, we introduce the RHNs to cancel the new gauge-gravity anomaly. The RHNs also participate in the mass generation of light neutrinos through the seesaw mechanism. 

\begin{table}
\centering
\vspace{0.5em}
\begin{tabular*}{\columnwidth}{l @{\extracolsep{\fill}} rrrr}
\hline
&${SU}(3)_{c}$&${SU}(2)_{L}$&${U}(1)_{Y}$&${U}(1)_{z}$\\
\cline{2-3}\cline{4-5}
&\multicolumn{2}{c}{representations}&\multicolumn{2}{c}{charges}\\
\hline\hline
$q_{L}$&$\mathbf 3$&$\mathbf 2$&$1/3$&$z_{q}$\\
$u_{R}$&$\mathbf 3$&$\mathbf 1$&$4/3$&$-z_{q}$\\
$d_{R}$&$\mathbf 3$&$\mathbf 1$&$-2/3$&$-z_{q}$\\~\\
$\ell_{L}$&$\mathbf 1$&$\mathbf 2$&$-1$&$0$\\
$e_{R}$&$\mathbf 1$&$\mathbf 1$&$-2$&$0$\\
$N_{R}$&$\mathbf 1$&$\mathbf 1$&0&$z_{N}$\\
$S$ & $\mathbf 1$ & $\mathbf 1$ & 0 & 0 \\~\\
$H$&$\mathbf 1$&$\mathbf 2$&$1$&$0$\\
$\varphi$&$\mathbf 1$&$\mathbf 1$&0&1\\
\hline
\end{tabular*}
\caption{The representations and charges for different particles in the GS model described in Sec.~\ref{subsec:gs}. Three fermions come in three generations. We assume generation-independent $U(1)_z$ charges for all fermions.}\label{tab:charges}
\end{table}

To realise the leptophobic nature of $Z'$, we assume the SM leptons ($\ell_L$ and $e_R$) do not carry any $U(1)_z$ charge. However, we assume both the left- and right-handed quarks to be charged under $U(1)_z$ to produce  $Z'$ at the LHC. There are many possible charge assignments one can think of. We assign a uniform $U(1)_z$ charge of $z_q$ to all the left-handed quarks and $-z_q$ to the right-handed ones. We introduce three RHNs all of which are equally charged under $U(1)_z$ with charge $Z_N$. We assume the SM Higgs doublet ($H$) to be chargeless under $U(1)_z$, and thus avoid tree-level $Z\leftrightarrow Z'$ mixing. We introduce a scalar flavon field $\varphi$ with unit $U(1)_z$ charge. For the ISM, we need three singlet chiral fermions, $S_i$, one for every generation, which are neutral under $U(1)_z$. The charge assignments of the fermion and scalar fields are summarised in Table~\ref{tab:charges}.

Six new nontrivial gauge anomalies arise from $U(1)_z$ with the corresponding group theory factors (for every generation):
\begin{enumerate}[label=~\Roman*.~]
    \item $[SU(3)_c]^2[U(1)_z]$: $\mathrm{tr}[\{\mc{T}^a,\mc{T}^b\}z]=12z_q$
    \item $[SU(3)_L]^2[U(1)_z]$: $\mathrm{tr}[\{T^i,T^j\}z]=6z_q$
    \item $[U(1)_Y]^2[U(1)_z]$: $\mathrm{tr}[Y^2z]=\dfrac{22}{3}z_q$
    \item $[U(1)_Y][U(1)_z]^2$: $\mathrm{tr}[Yz^2]=0$
    \item $[U(1)_z]^3$: $\mathrm{tr}[z^3]=12z_q^3 - z_N^3$
    \item $[R]^2[U(1)_z]$: $\mathrm{tr}[z]=12z_q - z_N$
\end{enumerate}
Note that the $[U(1)_Y][U(1)_z]^2$ anomaly vanishes identically for the charge assignments in Table~\ref{tab:charges}. 
The cancellation of the mixed gauge-gravity anomaly $[R]^2[U(1)_z]$ requires,
\begin{align}
\label{eq:mggrel}
12\, z_q -  z_N = 0.
\end{align}
Therefore, we remain with only one free charge in our model. To cancel the rest of the anomalies, we introduce a Peccei-Quinn (PQ) term for the pure $[U(1)_z]^3$ anomaly and a PQ and a generalised Chern-Simons (GCS) term for each of the mixed ones following Ref.~\cite{Ekstedt:2017tbo} (also see~\cite{Anastasopoulos:2006cz,Anastasopoulos:2008jt}),
\begin{align}
\label{eq:LagPQ}
\mathcal{L}_\text{PQ} = &\dfrac{1}{96\pi^2}
\lt(\dfrac{\Theta}{M}\rt)\varepsilon_{\mu\nu\rho\sigma}  \Big[g_z^2 \mc C_{zzz} F_z^{\mu\nu}
F_z^{\rho\sigma}+g_z g'\mc C_{zzy}  F_z^{\mu\nu} F_Y^{\rho\sigma}\nn\\
&+g'^2\mc C_{zyy}  F_Y^{\mu\nu} F_Y^{\rho\sigma}
+ g^2\mc D_{2}  \mathrm{tr}\left(F_W^{\mu\nu}
F_W^{\rho\sigma}\right)\nn\\
&+g_S^2\mc D_{3}  \mathrm{tr}\left(F_S^{\mu\nu}
F_S^{\rho\sigma}\right)\Big],\\
\mathcal{L}_\text{GCS}=&\frac{1}{48 \pi^2}\varepsilon_{\mu\nu\rho\sigma}\Big[\vphantom{\Omega_S^{\nu\rho\sigma}} g'^2 g_z \mc E_{zyy} B_Y^\mu B_z^\nu F_Y^{\rho\sigma}+g' g_z^2 \mc E_{zzy} B_Y^\mu B_z^\nu F_z^{\rho\sigma}\nn\\
&+ g^2 g_z \mc K_{2} B_z^\mu  \Omega_W^{\nu\rho\sigma}+g_S^2 g_z \mc K_{3} B_z^\mu\Omega_S^{\nu\rho\sigma}  \Big],
\end{align}
where
\begin{align}
\label{eq:omg}
\Omega_{G,W}^{\nu\rho\sigma} = &\frac{1}{3}\mathrm{tr}\Big[A_{G,W}^{\nu}\left(F_{G,W}^{\rho\sigma}-[A_{G,W}^{\rho},A_{G,W}^{\sigma}]\right)\nn\\
&+\left(\mathrm{cyclic~permutation}\right)\Big] \mbox{~with~} A_X=\{G,W\}.
\end{align}
In the above, $\{G^\mu, W^\mu, B_Y^\mu, B_z^\mu\}$, $\{g_S,g,g^\prime,g_z\}$, and $\{F_G^{\mu\nu},F_W^{\mu\nu},F_Y^{\mu\nu},F_z^{\mu\nu}\}$ are the gauge fields, gauge couplings and the field strength tensors associated with the $\{SU(3)_c, SU(2)_L,  U(1)_Y, U(1)_z\}$ groups, respectively and $\Theta$ is the axion. Under the $U(1)_z$ group, the axion and $B_z^\mu$ fields transform as $\Theta\to \Theta + Mg_z\theta_z$ and $B_z^\mu \to B_z^\mu - \partial^\mu\theta_z$, where $\theta_z$ is a scalar function of spacetime and $M$ is the $U(1)_z$-breaking scale. As the Abelian $U(1)_z$ breaks at the high scale $M$ through the St\"{u}ckelberg mechanism, it introduces a massive $Z'$ in the TeV scale.  The coefficients $\mc C, \mc D, \mc E$, and $\mc K$ are chosen such that the anomalies are cancelled. 

To make the Yukawa interactions $U(1)_z$ invariant, we consider the following higher-dimensional operators, 
\begin{align}
\mc{L} \supset -\lm_u\lt(\dfrac{\varphi}{\Lm}\rt)^{2z_q}\overline{q_L}\widetilde{H}u_R - \lm_d\lt(\dfrac{\varphi}{\Lm}\rt)^{2z_q}\overline{q_L} H d_R + h.c.\,,
\end{align}
where $H$ is the Higgs doublet, $\widetilde{H} = i \sigma_2 H^*$, and $\lm_{u,d}$ are coupling matrices. We obtain the SM Yukawa matrices by replacing $\varphi$ by its vacuum expectation value (VEV), $v_\varphi = \langle\varphi\rangle$. We make a benchmark choice of $z_q = 1/2$ [i.e., $z_N = 6$ from Eq.~\eqref{eq:mggrel}] so that the  $\varphi/\Lm$ factor only appears with integral powers in the Lagrangian (as it is needed to keep the theory local).  

The ISM can give us TeV-scale heavy sterile neutrinos with $\mathcal{O}(1)$ Yukawa couplings~\cite{Bhardwaj:2018lma, Das:2017pvt,Das:2012ze}. We can write the Lagrangian to generate the neutrino masses through the ISM~\cite{Mohapatra:1986aw,Mohapatra:1986bd} as, 
\begin{align}
\label{ISS_Lag}
\mathcal{L} \supset &\ Y_{ij}^{\nu}\lt(\dfrac{\varphi^\dag}{\Lm}\rt)^{z_N}\overline{L_i} \,\widetilde{H} \,N_{R_j} + M_{R_{ii}}\lt(\dfrac{\varphi}{\Lm}\rt)^{z_N}\overline{N_{R_i}} \,S_{L_i} \nn \\ 
& + \frac{1}{2} \mu_{ii} \overline {S^c_{L_i}} S_{L_i}+ h.c.
\end{align}
Here, $i,j$ are the generation indices. The $M_R$ and $\mu$ matrices can be considered diagonal if minimal flavour violation is assumed~\cite{PhysRevD.99.123508}. In that case, only the Dirac mass matrix $\mathbf{m_D}$ that arises from the term $Y_{ij}^{\nu}\lt(\varphi^\dag/\Lm\rt)^{z_N} \,\overline{L_i} \,\widetilde{H} \,N_{R_j}$ in the above Lagrangian causes the flavour violation. The mass matrix in the $\{ \nu_L^c, N_R, S_L^c\}$ basis can be written as
\begin{equation}
\mathbf M_\nu = \begin{pmatrix}\label{ISS_matrix}
\mathbf 0 && \mathbf{m_D} && \mathbf 0 \\
\mathbf{m_D}^T && \mathbf 0 && \mathbf{M_R} \\
\mathbf 0 && \mathbf {M_R}^T && \mathbf{\mu} \\
\end{pmatrix}.
\end{equation}  
One gets the light-neutrino masses by block-diagonalising the above matrix as
\begin{equation}\label{eq:mnu}
\mathbf m_\nu = \mathbf {m_D (M_R}^T)^{-1}\mathbf{\mu~ 
 M_R}^{-1}\mathbf{m_D}^T
\end{equation}
where the small scale $\mu$, defined as $\mu_{ij}=\mu\delta_{ij}$, is generally used as a measure of lepton number violation (the lepton number symmetry is restored in the $\mu\to 0$ limit). Note that the light neutrino masses are independent of the $(v_\varphi/\Lambda)^{2z_q}$ factor as it appears in both $\mathbf{m_D}$ and $\mathbf{M_R}$.
To find the sterile mass states, one has to go to the $\{N_R,\, S_L\}$ basis where one gets a $6\times6$ block matrix,
\begin{equation}
\mathbf{M_\nu^{6\times6}} = \begin{pmatrix}\label{6by6}
\mathbf 0 && \mathbf {M_R} \\
\mathbf {M_R}^T && \mathbf{\mu}  \\
\end{pmatrix}.
\end{equation}
Diagonalising it, we get the resulting pseudo-Dirac mass states~\cite{Dolan:2018qpy},
\begin{equation}\label{eq:splitting}
\mathbf{M_N} =  \frac{1}{2} \left( \mathbf{\mu} \pm \sqrt{\mathbf{\mu}^2 + 4\mathbf{M_R}^2}\right).
\end{equation} 

From the neutrino mass observations and the recent cosmological bound on the sum over neutrino masses ($\sum_i m_i\,\leq \,0.12$\,eV \cite{Planck:2018vyg}), assuming the normal ordering, we obtain the central values of the mass differences and the angles as~\cite{pdg_2020},
\begin{align*}
\Delta m_{21}^2 = 7.39~{\rm eV}^2, &\quad \Delta m_{32}^2 = 2.449~{\rm eV}^2\\
\sin^2(\theta_{12}) = 3.1 \times 10^{-1},&\quad \sin^2(\theta_{32}) = 4.5 \times 10^{-1},\\
\sin^2(\theta_{13}) = 2.246 \times 10^{-2}.\hspace{-2cm}&
\end{align*}
 In order to obtain the complete Yukawa texture of the Dirac coupling ($Y_{ij}^\nu$) one can use the Casas-Ibarra formalism~\cite{Casas:2001sr,Dolan:2018qpy,Mukherjee:2022fjm}\footnote{The rotational mixing angle choices in the $R$ matrix has been kept to be $x \rightarrow \pi/4,\, y \rightarrow \pi/2,\, z \rightarrow \pi/3$. The orders of magnitude of the Yukawa coupling and the light-heavy mixing are equally sensitive to any other set of these values.} to get,
 \begin{equation}\label{eq:CI}
\mathbf{Y^{\nu}} = \frac{1}{v} \,\mathbf{U} \,\mathbf{m_{n}}^{1/2} \,\mathbf{R} \, \mathbf{\mu}^{-1/2}\,\mathbf{M_{N}}^T\,,
\end{equation}
where $\mathbf{U}$ is the neutrino mixing matrix, $\mathbf{R}$ is a rotation matrix, $v$ is the Higgs VEV, and $\mathbf{m_n} \equiv \text{diag}(m_{1},m_{2},m_{3})$ and $\mathbf{M_{N}} \equiv \text{diag}(M_{1},M_{2},M_{3})$\footnote{One can also take a democratic structure for $M_N$ with the choice $M_1 \approx M_2 \approx M_3$, which is a feature of the ISM itself due to the scale $\mu$ creating such near degeneracy.} are $3\times3$ mass matrices.
 
With the best fit central values of the neutrino oscillation parameters and the low-energy $CP$-violating sources switched off (by setting the Dirac and Majorana phases in the neutrino mixing matrix to zero, as their presence does not affect our results), we get  
\begin{equation*} \label{eq:yukawa}
  \mathbf{Y^\nu} =  \left(
\begin{array}{rrr}
 0.004 & -0.013 & ~~0.054 \\
 -0.013 & 0.027 & 0.081 \\
 -0.053 & 0.007 & 0.043 \\
\end{array}
\right).
\end{equation*}
To obtain this, we have assumed $v_\varphi/\Lm \sim 1$. (Note that even though there are different scales in this model, they do not have an unnatural hierarchy. The St\"{u}kelberg scale $M$, the Froggatt-Nielsen scale $\Lm$, the Majorana mass scale $M_R$, and the VEV $v_{\varphi}$ are all in the TeV range.) With the above Yukawa matrix and our choice of scales, the light-heavy neutrino mixing stands in the order of $10^{-3}$ or less and hence, our choice of parameters remains safe from the current LFV bounds and, at the same time, does not lead to any displaced vertex in the $N_R$ decay at the LHC.

\begin{table}
\centering
\vspace{0.5em}
\begin{tabular*}{\columnwidth}{l @{\extracolsep{\fill}} rrrrr}
\hline
&${SU}(3)_{c}$&${SU}(2)_{L}$ &${U}(1)_{Y}$ & $U(1)_\psi$ & $U(1)_\chi$ \\
\cline{2-3}\cline{4-6}
&\multicolumn{2}{c}{representations}& $Y$ & $Q_\psi$ & $Q_\chi$ \\
\hline\hline
$q_{L}$&$\mathbf 3$&$\mathbf 2$&$1/3$ & $1/2\sqrt{6}$ & $-1/2\sqrt{10}$\\
$u_{R}$&$\mathbf 3$&$\mathbf 1$ & $4/3$ & $-1/2\sqrt{6}$ & $1/2\sqrt{10}$\\
$d_{R}$&$\mathbf 3$&$\mathbf 1$ & $-2/3$ & $-1/2\sqrt{6}$ & $-3/2\sqrt{10}$\\~\\
$\ell_{L}$&$\mathbf 1$&$\mathbf 2$& $-1$ & $1/2\sqrt{6}$ & $3/2\sqrt{10}$\\
$e_{R}$&$\mathbf 1$&$\mathbf 1$& $-2$ & $-1/2\sqrt{6}$ & $1/2\sqrt{10}$ \\
$N_{R}$&$\mathbf 1$&$\mathbf 1$ & 0 & $-1/2\sqrt{6}$ & $5/2\sqrt{10}$\\
\hline
\end{tabular*}
\caption{The representations and charges for the SM fields and the RHNs in the standard embedding of the $\mathbf{27}$ representation of $E_6$~\cite{Leroux:2001fx}. These charges are generation independent.}
\label{tab:gutcharges}
\end{table}

\subsection{Leptophobic $Z^\prime$ in a GUT model}\label{subsec:gut}\noindent
Generally, there is no leptophobia in 
conventional GUT models since the fermion couplings are determined by their embeddings in the gauge group. However, some distinct regions of the parameter space can show leptophobia through kinetic 
mixing~\cite{Lopez:1996ta,Rizzo:1998ut,Leroux:2001fx}. For our purpose, we assume that the $Z'$ arises from the breaking of the $E_6$ group to $SU(2) \times U(1)_Y\times U(1)_z$ (we provide a brief overview here---more details are found in Refs.~\cite{Rizzo:1998ut,Leroux:2001fx}). The symmetry-breaking chain goes as follows:
\begin{align}
    E_6 ~&\to~ SO(10)\times U(1)_\chi \nn \\
    &\to~ SU(5) \times U(1)_\chi \times U(1)_\psi \nn \\
    &\to~ SU(2)_L \times U(1)_Y \times U(1)_z \nn \\
    &\to~ SU(2)_L \times U(1)_Y.
\end{align}
We identify $U(1)_z$ as a linear combination of $U(1)_\psi$ and $U(1)_\chi$, with $Q_z = Q_\psi\cos\theta - Q_\chi\sin\theta$, where $\theta$ is the $E_6$ mixing angle and $Q_{\psi,\chi}$ are the quantum numbers of the particles in the fundamental $\mathbf{27}$ representation of $E_6$. The $\mathbf{27}$ representation decomposes further to a $\mathbf{16}+\mathbf{10}+\mathbf{1}$ under $SO(10)$. We assume the standard embedding where all the SM particles, along with a RHN, are put in the $\mathbf{16}$. There is no solution of $\theta$ for which leptophobia can be achieved if the $Z'$-couplings to fermions are proportional to $Q_z$. However, a kinetic mixing term between $U(1)_Y$ and $U(1)_z$ of the form,
\begin{align}
 - \dfrac{1}{2}\sin\alpha~ \widetilde{B}^{\mu\nu}\widetilde{Z}^\prime_{\mu\nu}\,,
\end{align}
can lead to leptophobia. Here, $\sin\alpha$ is the kinetic mixing parameter. The kinetic mixing can be rotated away by the following transformations,
\begin{align}
\widetilde{B}_\mu = B_\mu - \tan\alpha~Z^\prime_{\mu};~~~\widetilde{Z}^\prime_\mu = \dfrac{Z^\prime_\mu}{\cos\alpha}\,,
\end{align}
where $\widetilde{B}_\mu$ (${B}_\mu$) and $\widetilde{Z}^\prime_\mu$ (${Z}^\prime_\mu$) are the $U(1)_Y$ and $U(1)_z$ gauge fields before (after) the above rotation, respectively. The corresponding gauge couplings are related as, $g' = \widetilde{g}^\prime$ and $g_z = \widetilde{g}_z/\cos\alpha$.
Finally, the normalised $Z'$ interaction to a fermion $f$ can be written as
\begin{align}
\mc{L} \supset - g_z\lt(\sqrt{\dfrac{5}{3}}Q_z + \delta \dfrac{Y}{2}\rt)\overline{f}\gamma^\mu f\ Z^\prime_\mu
\end{align}
In Table~\ref{tab:gutcharges}, we summarise the charges of the SM fields along with the RHN in the standard embedding. There are additional fields present in the $\mathbf{27}$ representation of $E_6$. However, we do not list them in Table~\ref{tab:gutcharges} assuming they are heavier than the $Z'$. Therefore, they do not contribute to the $Z'$ BRs and are irrelevant to our results.

The above coupling depends on two free parameters, $\theta$ and $\delta$. One can, in principle, make two couplings of the $Z'$ vanish. For leptophobia, we demand that the $Z'$ couplings to $\ell_L$ and $e_R$ vanish simultaneously. Of the six possible embeddings~\cite{Rizzo:1998ut,Leroux:2001fx}, one possible leptophobic solution is $\theta = \tan^{-1}\sqrt{3/5}$ and $\delta=-1/3$. Note that it is only the SM leptons to which the $Z'$ has no couplings. However, it still couples to the RHNs naturally present in the GUT framework. The RHNs can give masses to the light neutrinos through the seesaw mechanism. We, however, do not repeat a similar discussion on neutrino mass generation in this context.

\begin{figure*}
\captionsetup[subfigure]{labelformat=empty}
\subfloat[\quad\quad\quad(a)]{\includegraphics[width=0.95\columnwidth]{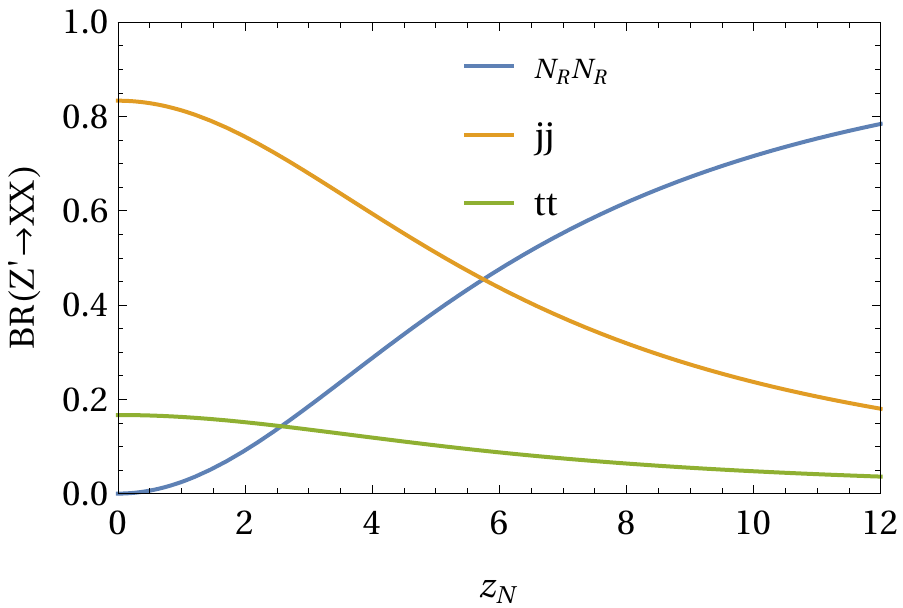}\label{fig:BRZpNN}}\hspace{1cm}
\subfloat[\quad\quad\quad(b)]{\includegraphics[width=0.95\columnwidth]{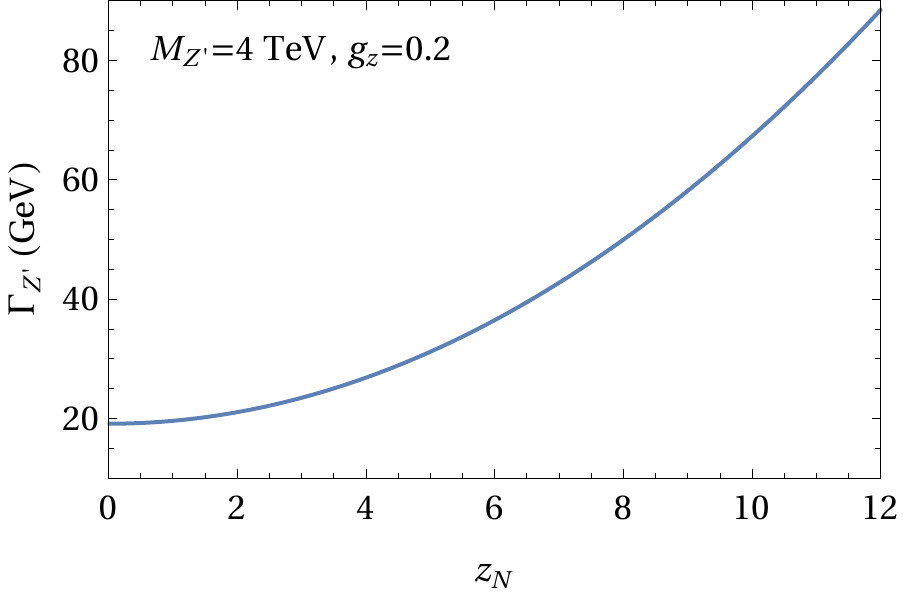}\label{fig:TWZp}}
\caption{(a) The branching ratios of the $Z^\prime\to NN,jj,tt$ decay modes and (b) the total decay width of $Z^\prime$ as  functions of $z_N$. The $jj$ mode includes the $b\bar b$ final state. For these plots, we have set $M_{Z^\prime}=4$ TeV, $M_{N}=0.5$ TeV, $g_z=0.1$, and $z_u=z_d=1$. }
\label{fig:BRTWZp}
\end{figure*} 
\section{$Z^\prime$ decays and bounds}
\label{sec:zpbound}
\noindent
The leptophobic models with the $Z^\prime\to N_RN_R$ decay can be parametrised in a simple manner. For our phenomenological analysis,
we consider the following Lagrangian for a leptophobic $Z^\prime$ to make our presentation model independent,
\begin{align}
\label{eq:modlag}
\mc{L} \supset&\ \dfrac{g_z}{2}\big(z_{u_L} \bar{u}^i_L\gm^\mu u^i_L + z_{u_R} \bar{u}^i_R\gm^\mu u^i_R+ z_{d_L} \bar{d}^i_L\gm^\mu d^i_L  \nn\\
&\quad\quad+ z_{d_R} \bar{d}^i_R\gm^\mu d^i_R + z_{N} \bar{N}_R\gm^\mu N_R\big) Z^{\prime}_\mu\ ,
\end{align}
where $z_{u_{L/R}}$, $z_{d_{L/R}}$ and $z_N$ are the $U(1)_z$ charges of left/right-handed up-type, down-type quarks and the RHNs, respectively. The new $U(1)_z$ gauge coupling is denoted by $g_z$. In general, in an anomaly-free $U(1)$ extension, the right- and left-handed projections of a fermion have different $U(1)$ charges (like the hypercharges in the SM). They can be generation dependent too. Here, we simply assume generation-independent $U(1)_z$ charges for all fermions. For a collider analysis, we can reduce the number of free parameters further. The production cross section $\sg(pp\to Z')$ and the partial decay widths of the $Z^\prime\to qq$ decays are proportional to the sum of the square of the left and right couplings of the quarks. Therefore, it is possible to assume a single effective coupling for a given quark type as long as we do not use any asymmetry observable sensitive to the left and right couplings separately. Hence, we can simplify the above Lagrangian further as
\begin{align}
\label{eq:modindlag}
\mc{L} \supset \dfrac{g_z}{2}(z_u \bar{u}_i\gm^\mu u_i + z_d \bar{d}_i\gm^\mu d_i + z_{N} \bar{N}_R\gm^\mu N_R) Z^{\prime}_\mu\,,
\end{align}
with $z_{q}^2 = z_{q_L}^2+z_{q_R}^2$. Since the up- and down-type quarks (in a given generation) have different parton distribution functions (PDFs), we need to keep two separate free charges, one for the up-type quarks ($z_u$) and the other for down-type quarks ($z_d$), as different choices of $z_u$ and $z_d$ can change the kinematic distributions. With these simplifications, we now have a total of six free parameters in the model---the two masses, $M_{Z^\prime}$ and $M_{N_R}$, the $U(1)_z$ gauge coupling $g_z$, and the three charges $z_{u,d,N}$.

The tree-level partial decay widths of $Z^\prime$ to a quark pair and a RHN pair are given by the following expressions,
\begin{align}
\Gm(Z^\prime\to qq) &= \dfrac{g_z^2 z_q^2}{16\pi}M_{Z^\prime}\lt(1+\dfrac{2M_q^2}{M_{Z^\prime}^2}\rt) \lt(1-\dfrac{4M_q^2}{M_{Z^\prime}^2}\rt)^{1/2}\nn\\
\Gm(Z^\prime\to N_RN_R) &= \dfrac{g_z^2 z_N^2}{96\pi}M_{Z^\prime}\lt(1-\dfrac{4M_{N_R}^2}{M_{Z^\prime}^2}\rt)^{3/2}.
\end{align}
We show the BRs and the total width of $Z^\prime$ as functions of $z_N$ (while fixing the other free parameters fixed at some benchmark values) in Fig.~\ref{fig:BRTWZp}. As we increase $z_N$ keeping $z_u$ and $z_d$ fixed, the $Z^\prime\to N_RN_R$ BR increases. Hence, the importance of searching for $Z^\prime$ in the di-RHN mode grows in the models with comparatively larger $z_N$. The total width plot confirms the validity of the narrow-width approximation in our case.

In our leptophobic $U(1)_z$ models, the $Z^\prime$ would be produced at the LHC through the quark-antiquark fusion processes.
We parametrise the production cross section of $Z^\prime$ as
\begin{equation}
\sg(pp\to Z^\prime) = K_{QCD}\times \dfrac{g_z^2}{4}\lt[z_u^2\sg_u(M_{Z^\prime})+z_d^2\sg(M_{Z^\prime})\rt],
\end{equation}
where $\sg_u(M_{Z^\prime})$ and $\sg_d(M_{Z^\prime})$ parametrise the respective contributions from the up-type (except the top) and down-type quarks. The higher-order QCD corrections to the production cross section are factored in $K_{QCD}$. We use a constant $K_{QCD}$ of $1.3$ for all values of $M_{Z^\prime}$~\cite{Eichten:1984eu}.

\begin{figure}
\captionsetup[subfigure]{labelformat=empty}
\subfloat[(a)]{\includegraphics[height=6.5cm,width=7.5cm]{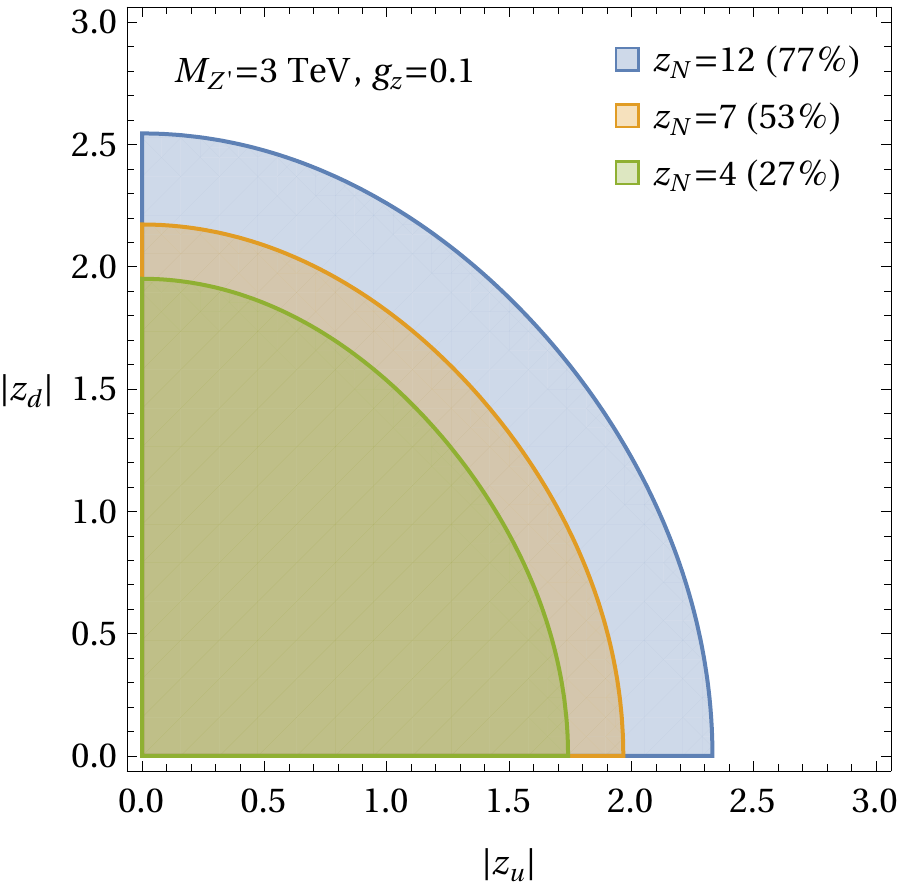}\label{fig:zuzdzn}}\\
\subfloat[(b)]{\includegraphics[height=6.5cm,width=7.5cm]{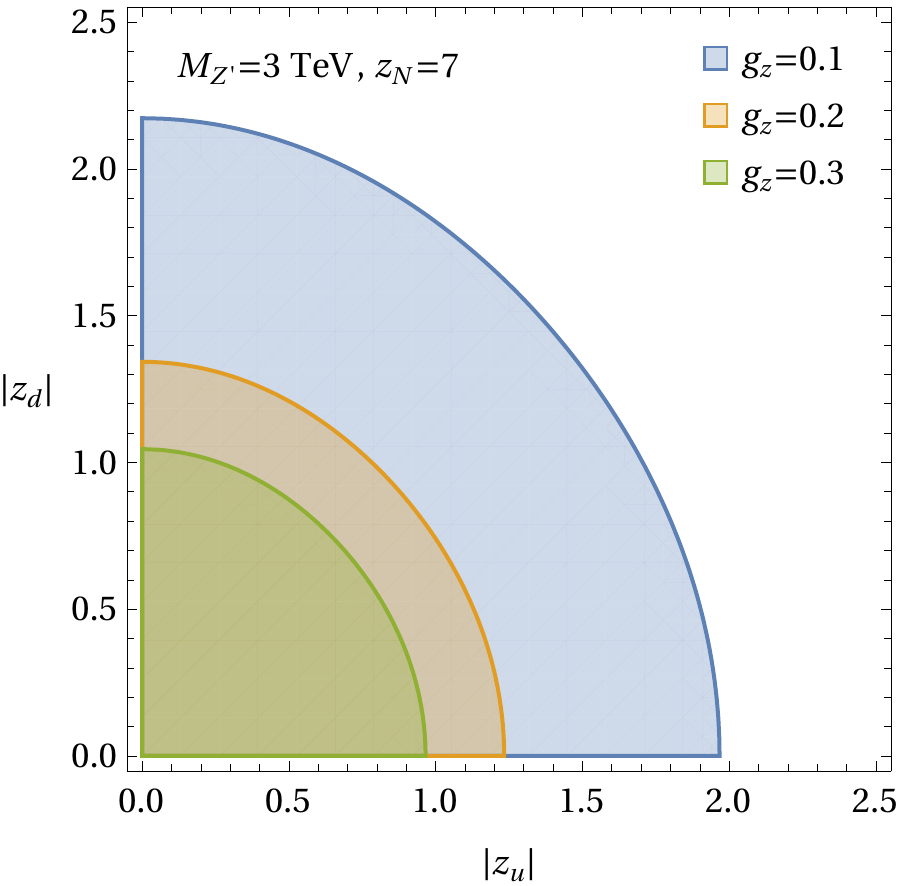}\label{fig:zuzdgx}}\\
\subfloat[(c)]{\includegraphics[height=6.5cm,width=7.5cm]{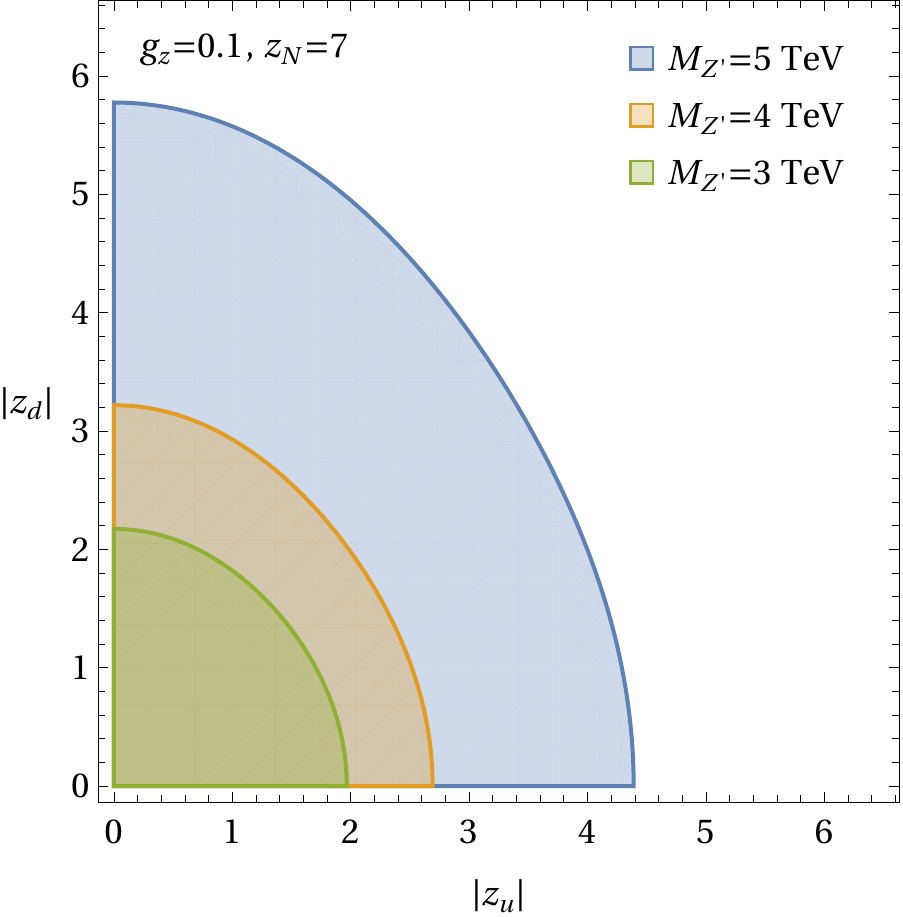}\label{fig:zuzdmzp}}
\caption{Allowed regions in the $z_u-z_d$ plane from the dijet resonance search data~\cite{ATLAS:2019fgd}. The $Z^\prime\to N_RN_R$ branching ratio is roughly $27$\% in (a), $53$\% in (b), and $77$\% in (c).}
\label{fig:zuzdalreg}
\end{figure}

The leptophobic $Z'$ would decay to both fermionic ($jj$, $tt$ and $N_RN_R$) and bosonic ($WW$ and $ZH$) final states. Since, the two bosonic decays of $Z^\prime$ are $Z\leftrightarrow Z^\prime$ mixing-angle suppressed, the major constraints on the free parameters come from the dijet resonance searches at the LHC. The $tt$ resonance search data give less restricted bounds than the dijet data due to less sensitivity. In Fig.~\ref{fig:zuzdalreg}, we recast the latest ATLAS dijet resonance search data~\cite{ATLAS:2019fgd} to obtain the allowed regions in the $z_u-z_d$ plane. We choose three benchmark values of $z_N$ for which  BR($Z^\prime\to N_RN_R$) is about $25$\%, $50$\%, and $75$\%. The open regions in the $z_u-z_d$ plane are elliptic in shape since the PDFs of the up and down quarks in proton are different. 

\section{Pair production of $N_R$}
\label{sec:sigpro}
\noindent 
If the RHNs are highly charged under the $U(1)_z$, the $Z^\prime\to N_RN_R$ mode can have high BR. In that case, the dijet constraints would relax, making the $pp\to Z^\prime\to N_RN_R$ channel a promising new channel for the discovery of $Z'$ at the HL-LHC.
The RHNs can mix with the SM neutrinos and decay to the SM states: $W^\pm\ell^\mp,~Z(\nu+\bar{\nu}),~H(\nu+\bar{\nu})$. The Goldstone boson equivalence theorem tells us that the $N_R\to W^\pm\ell^\mp$ decay has $50$\% BR whereas $Z(\nu+\bar{\nu})$ and $H(\nu+\bar{\nu})$ have $25$\% each in the large-$M_{N_R}$ limit. The proportions would slightly change because of the number of the RHNs, their mixing effects and the presence of Majorana phases. For our analysis, however, we assume a fixed $50$\% BR for the $N_R\to W^\pm\ell^\mp$ mode for simplicity. This means about $25$\%
of the $N_R$ pairs produced at the LHC via $Z^\prime$ decay would show our desired signature, i.e., a same-flavour opposite-sign lepton pair plus two $W$-like fatjets (Fig.~\ref{fig:FD}). If, however, the RHNs were Majorana fermions, then a neutrino pair would show the OSDL or SSDL signatures with $12.5$\% probability. Hence a charge-agnostic selection criterion in the experiment, as is often the case,  will not let us distinguish the nature of the RHNs. Nonobservation of any same-sign dilepton signature along with the opposite-sign dilepton events can hint towards the inverse seesaw mechanism of neutrino mass generation.  

Apart from the OSDL channel, the $pp\to Z^\prime\to N_RN_R$ process would lead to more interesting final states. In terms of the decay modes of RHN, we can have two types of channels namely the symmetric, when both the $N_R$'s decay to the same final state, and the asymmetric, when they decay to different final states. One can also categorise these channels in terms of the number of charged leptons in the final state. We discuss some of these channels below.

\vspace{0.25cm}
\noindent
\textbf{Monolepton:} The monolepton final states come from some asymmetric decays of $N_R$,
\begin{eqnarray}
pp \to Z^\prime\to N_RN_R \to \left\{\begin{array}{l}
(W^\pm_h\ell^\mp)(Z_h\nu) \\
(W^\pm_h\ell^\mp)(H_h\nu)
\end{array}\right\}.
\end{eqnarray}
Here, $W_h$, $Z_h$ and $H_h$ stand for the hadronic decays of the gauge and Higgs bosons. The full $Z^\prime$ reconstruction in this channel is not possible due to the missing energy (though, one RHN can be reconstructed). Moreover, the SM background is also huge. Possibly because of these reasons, the prospects of this channel at the HL-LHC are not available in the literature (to the best of our knowledge). In terms of BRs, this channel has a higher rate than the multilepton channels discussed below. Moreover, one could use the jet-substructure techniques to isolate the signal from the background as the signal has two fatjets from the hadronic decays of the SM gauge bosons. The dominant background to this process is $pp\to W +jets\to \ell\nu+jets$ where the lepton neutrino pair comes from a resonant $W$ decay, whereas, in the signal, the lepton and the neutrino come from two different heavy RHNs. Hence, the $\ell\nu$ pair is kinematically distinct in the signal and the background. This feature can be use to tame the huge background. Therefore, it would be interesting to obtain the projection of the monolepton channel for the HL-LHC using machine-learning techniques. 

\vspace{0.25cm}
\noindent
\textbf{Dilepton:} The dilepton final states arise from some symmetric decays of a $N_R$ pair,
\begin{equation}
pp \to Z^\prime\to N_RN_R \to (W^\pm_h\ell^\mp)(W^\pm_h\ell^\mp).
\end{equation}
The dilepton final state can also come from the leptonic decays of the $Z$ boson:
$N_RN_R\to (Z_\ell\nu)(Z_h\nu)+(Z_\ell\nu)(H_h\nu)$. However, we neglect this contribution mainly because of two reasons. First, the contribution of this channel is insignificant due to the small $Z\to \ell\ell$ branching and the fact that it cannot be fully reconstructed due to the missing energy. Second, to suppress the huge Drell-Yan dilepton background, a $Z$-veto cut would be necessary, which would also eliminate a part of this small signal.

If the RHNs were Majorana fermions, as it would have been the case in the standard type-I seesaw models, they could violate the lepton number by two units and produce a unique SSDL signature~\cite{Cox:2017eme,Das:2017flq,Das:2017deo}.  
The SSDL channel is almost free from the background and is commonly expected to give better sensitivity. 

\vspace{0.25cm}
\noindent
\textbf{Trilepton:} Both symmetric and asymmetric decays of a $N_R$ pair can lead to the trilepton signature,
\begin{eqnarray}
pp \to Z^\prime\to N_RN_R \to\left\{\begin{array}{l}
 (W^\pm_\ell\ell^\mp)(W^\pm_h\ell^\mp) \\
(W^\pm_h\ell^\mp)(Z_\ell\nu) 
\end{array}\right\}.
\end{eqnarray}
It has been investigated in Refs.~\cite{Kang:2015uoc,Accomando:2017qcs} where it was shown that this channel has good sensitivity over the SM background. A four-lepton final state is also possible from the $N_RN_R$ decays. It has been investigated in Ref.~\cite{Huitu:2008gf} in the context of a $\mathrm{U}(1)_{B-L}$ model. 

\vspace{0.25cm}
\noindent
\textbf{Displaced vertex:}
If the decay widths of the RHNs are very small (which happens when the tiny light-heavy neutrino mixing angle dictates the decays of the RHNs), they become long-lived and might lead to displaced vertices~\cite{Deppisch:2019kvs,Das:2019fee,Chiang:2019ajm}. When the decaying particles are highly boosted, their lifetimes in the lab frame are enhanced by the time-dilation effect. This is usually the case for the lighter RHNs. The pair production of light RHNs from a $Z^\prime$ in the $U(1)_{B-L}$ models has been investigated in Ref.~\cite{Padhan:2022fak}. Since the RHNs are TeV-scale particles in our model, they are not very boosted and therefore, do not show the displaced vertex signature.

\section{The signal and background processes}
\label{sec:sigback}
\noindent
The public packages we use for our analysis are as follows.
We obtain the Universal FeynRules Output~\cite{Degrande:2011ua} model files for 
the Lagrangian in Eq.~\eqref{eq:modlag} with \textsc{FeynRules}~\cite{Alloul:2013bka}. We use \textsc{MadGraph5}~\cite{Alwall:2014hca} to generate the signal and background events at the leading order using \textsc{NNPDF2.3LO} PDFs~\cite{Ball:2012cx}. For event generation, we use the default dynamical renormalisation and factorisation scales in \textsc{MadGraph5}. Events are first passed through \textsc{Pythia8}~\cite{Sjostrand:2014zea} for showering and hadronisation and then subsequently through \textsc{Delphes}~\cite{deFavereau:2013fsa} for simulating the detector environment. Jets are formed from the tower objects using the anti-$k_t$ jet clustering algorithm~\cite{Cacciari:2008gp} in \textsc{FastJet}~\cite{Cacciari:2011ma}.
In our analysis, jets with two different jet-radii ($R$'s) have been used~\cite{Bhaskar:2021gsy}; the AK4-jets with $R=0.4$ (denoted as `$j$') and the AK8-fatjets with $R=0.8$ (written as `$J$'). 

\subsection{The signal}
\noindent
As mentioned earlier, we are interested in  a same-flavour opposite-sign lepton (muon) pair and $W$-like fatjets in the final state,
\begin{align}
p p \to Z^\prime &\to N_R N_R  \to \mu^{+}\mu^{-} + 2J.
\end{align}
A representative Feynman diagram is shown in Fig.~\ref{fig:FD}.
For demonstration, we assume that out of the three generations of RHNs, only one that couples with the muon is lighter than $M_{Z^\prime}/2$ so that it can be produced from the $Z^\prime\to N_RN_R$ decay. 
Moreover, when it decays, it produces a muon through the  $N_R\to W^\pm\m^\mp$ decays.
The choice of muon is motivated by the fact that the muon-detection efficiency is high at the LHC.

\begin{table}[!t]
\centering{\linespread{2}
\begin{tabular*}{\columnwidth}{l @{\extracolsep{\fill}} lrc }
\hline
\multicolumn{2}{l}{Background } & $\sg$ & QCD\\ 
\multicolumn{2}{l}{processes}&(pb)&order\\\hline\hline
\multirow{2}{*}{$V +$ jets~ \cite{Catani:2009sm,Balossini:2009sa}}   & $Z +$ jets  &  $6.33 \times 10^4$& NNLO \\
   & $W+$ jets  &  $1.95 \times 10^5$& NLO
\\ \hline
$tt$~\cite{Muselli:2015kba}  & $tt +$ jets  & $988.57$ & N$^3$LO\\ \hline
Single $t$~\cite{Kidonakis:2015nna}  & $tW$  &  $83.10$ & N$^2$LO \\  \hline
\multirow{3}{*}{$VV +$ jets~\cite{Campbell:2011bn}}   & $WW +$ jets  & $124.31$& NLO\\ 
                  & $WZ +$ jets  & $51.82$ & NLO\\ 
                   & $ZZ +$ jets  &  $17.72$ & NLO\\ \cline{1-4}
\multirow{2}{*}{$ttV$~\cite{Kulesza:2018tqz}} & $ttZ$  &  $1.05$ &NLO+NNLL \\ 
                   & $ttW$  & $0.65$& NLO+NNLL \\ \hline
\end{tabular*}}
\caption{The SM background processes considered in our analysis and their higher-order QCD cross sections~\cite{Bhaskar:2020gkk}. The corresponding QCD orders are shown in the last column. These cross sections are used to compute the constant QCD $K$-factors that we multiply with the tree-level cross sections obtained from \textsc{Madgraph} to incorporate higher-order effects.}
\label{tab:BGxsec}
\end{table}
\begin{table*}
\centering{\linespread{3}
\begin{tabular*}{\textwidth}{l @{\extracolsep{\fill}} r r r r r r r}
\hline
Selection cut & Signal & $Z+\textrm{jets}$ & $tt+\textrm{jets}$ & $tW+\textrm{jets}$ & $WW+\textrm{jets}$ & $ttW$ & $ttZ$ \\
\hline\hline
Generation level (including $K$ factors) & 252
 & $3.3\times 10^5$ & $4.4\times 10^5$ & $1.8\times 10^4$ & 9458 & 877 & 327 \\
Number of muons $=2$ (any charge) & 179
 & $2.2\times 10^5$ & $2.2\times 10^5$ & $1.1\times 10^4$ & 7820 & 480 & 160\\
Number of $b$-jet~$=0$ (AK4 jets)& 176
 & $2.2\times 10^5$ & $1.4\times 10^5$ & $1.0\times 10^4$ & 7780 & 323 & 108 \\
Selection cut $\mc C_1$& 171
 & $1.6\times 10^5$ & $8.1\times 10^4$ & 6840 & 6918 & 323 & 62 \\
Selection cut $\mc C_2$ & 169
 & $1.2\times 10^5$ & $7.9\times 10^4$ & 6630 & 6530 & 214 & 61 \\
Selection cut $\mc C_3$ & 121
 & $2.6\times 10^4$ & $1.3\times 10^4$ & 1148 & 1633  & 61 & 16 \\
Selection cut $\mc C_4$ & 115
 & 5937 & 1692 & 198 & 444 & 7 & 1 \\
Selection cut $\mc C_6$ & 109  & 212 & 23 & 1  & 12 & $<1$ & $< 1$ \\
\hline  
\end{tabular*}}
\caption{Number of signal and background events surviving the selection cuts defined in the text. These numbers are computed for the $\sqrt{s}=14$~TeV LHC with $\mc{L}=3000$~fb$^{-1}$. In the first row, the numbers of generation-level events are estimated by applying the generation-level cuts at the parton level. We use the benchmark point $M_{Z^\prime}=3500$ GeV, $M_{N_R}=1000$~GeV and $g_z=0.1$ to get these numbers. The selection cut $\mc{C}_5$ is not applicable for this benchmark choice.
}\label{tab:cutflow}
\end{table*}
\begin{figure*}
\captionsetup[subfigure]{labelformat=empty}
\subfloat[(a)]{\includegraphics[width=0.45\textwidth]{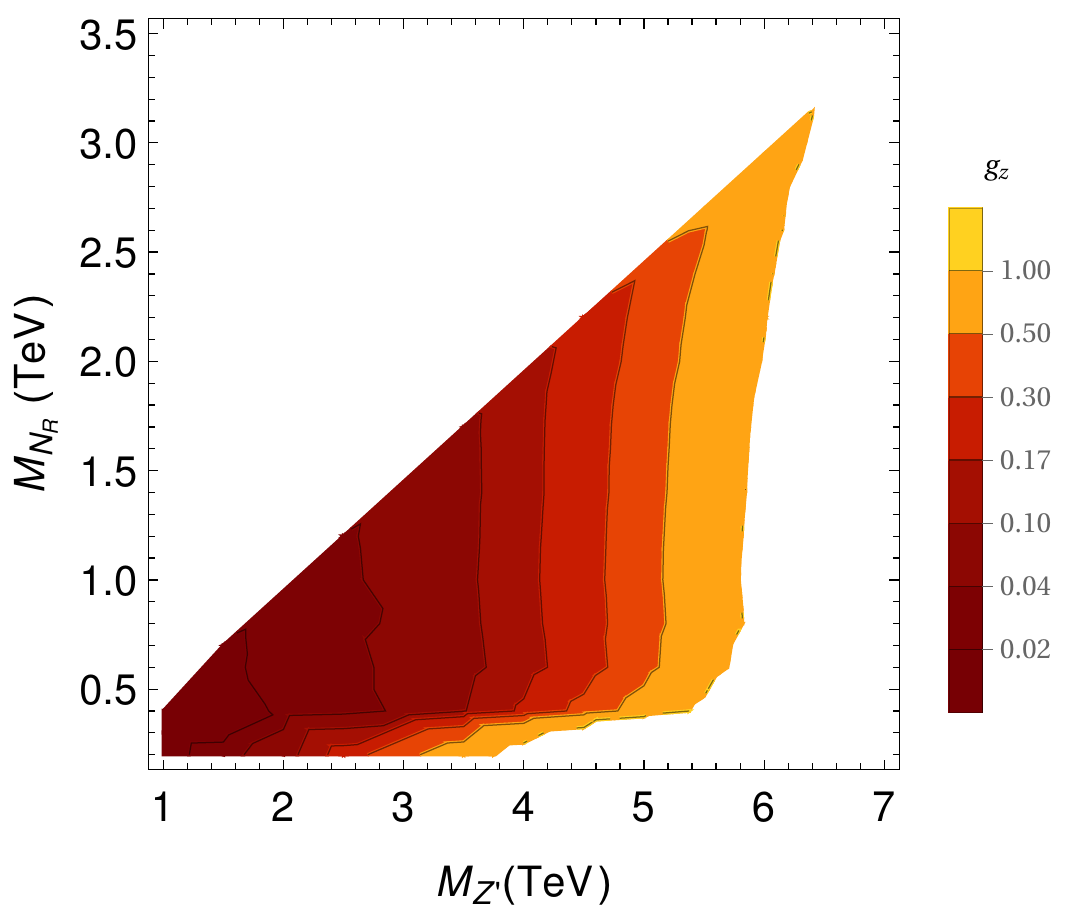}\label{fig:5sig}}\hspace{1cm}
\subfloat[(b)]{\includegraphics[width=0.45\textwidth]{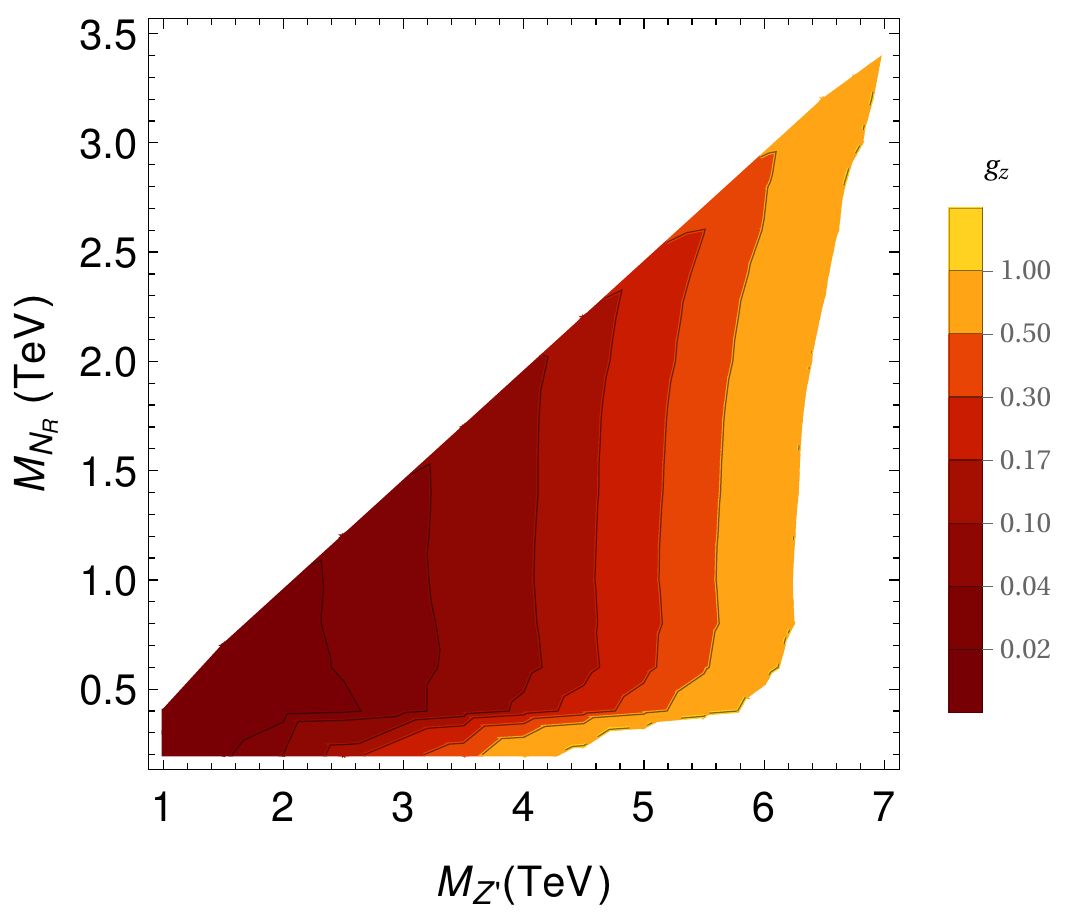}\label{fig:2sig}}
\caption{$M_{Z^\prime}-M_{N_R}$ plot demonstrating regions (a) for discovery with 5$\sigma$ significance (b) for exclusion with 2$\sigma$ significance. The contours in the left (right) plot correspond to different values of $g_z$.}
\label{fig:signi}
\end{figure*}

\subsection{Dominant background processes}
\noindent
In the SM, the dominant sources of dilepton and jets are as follows:

\begin{itemize}
    \item \emph{$Z+jets$:} The largest source of dileptons in the SM is
the Drell-Yan process, $pp \rightarrow Z/\gamma^* \rightarrow \ell^+\ell^-$. We
simulated it by matching up to two extra partons. The extra QCD jets can mimic the $W$-like fatjets. This is the major background for our signal. However, it can be tamed with a strong $Z$-mass-veto cut. Another monoboson process with large cross section is $pp\to W\to \ell\nu$ which gives one lepton. The second lepton can come from a jet faking as a lepton. However, due to the small mistagging efficiency ($\sim 10^{-4}$~\cite{Curtin:2013zua}), this process does not play any significant role.

\item
\emph{$W+jets$:} This process produces one lepton when the $W$ decays leptonically. However, similar to the $Z+jets$ case, a fake lepton can come from the jets. We consider it despite the mistagging efficiency as its cross section is large, it is of the order of $10^5$ pb. However, its contribution to the final background is negligible. (For the same-sign dilepton signal, this process is one of the major backgrounds.) We include up to three additional jets while generating the matched sample. 

\item \emph{$tt+jets$:}
The pair production of top quarks also acts as a source of high-$p_T$
dileptons when both decay leptonically. Since the QCD jets can mimic $W$-like fatjets, this process can lead to a similar final state as our signal. It forms one of the major sources of the background. We  generate this process by matching up to two jets. 

\item \emph{$tW+jets$:}
Single top process such as $pp\to tW$ also contributes to the background when both  the top and $W$ decay leptonically and the QCD jets mimic the  $W$-like fatjets. We generate it in a five-flavour scheme by matching the hard process with up to two additional jets. This process contributes significantly to the background.

\item \emph{$VV+jets$:} Two same-flavour leptons can also come from the following background processes: $W_\ell W_\ell$, $W_h Z_\ell$, $Z_\ell Z_h$ and $Z_\ell H_h$ (the subscripts ``$\ell$'' and ``$h$'' denote the leptonic and hadronic decays, respectively). A $Z$-mass veto can effectively control the processes involving the $Z\to \ell\ell$ decay. The $W_\ell W_\ell$ process turns out to be the top contributor among all the diboson processes. We generate the events of these processes by matching them with up to two extra jets. However, as we will see in the next section, the  total contribution of these processes in the final background after applying all the cuts is not significant.

\item \emph{ttV:}
The associated production of a massive vector boson with a top-quark pair can also act as a background. Depending on the decay modes, the processes, $t_\ell t_\ell Z_h$, $t_\ell t_\ell W_h$, $t_h t_\ell W_\ell$ and $t_\ell t_\ell H_h$ can contribute to the background. These are minor backgrounds and hence, we generate these without adding any extra jets.

\end{itemize}

 The cross sections of the background processes at the highest order in QCD available in the literature are listed in Table~\ref{tab:BGxsec}. We include the higher-order cross sections in our analysis through $K$ factors.

\begin{figure*}
\captionsetup[subfigure]{labelformat=empty}
\subfloat[\quad\quad\quad(a)]{\includegraphics[height=6.5cm,width=7.5cm]{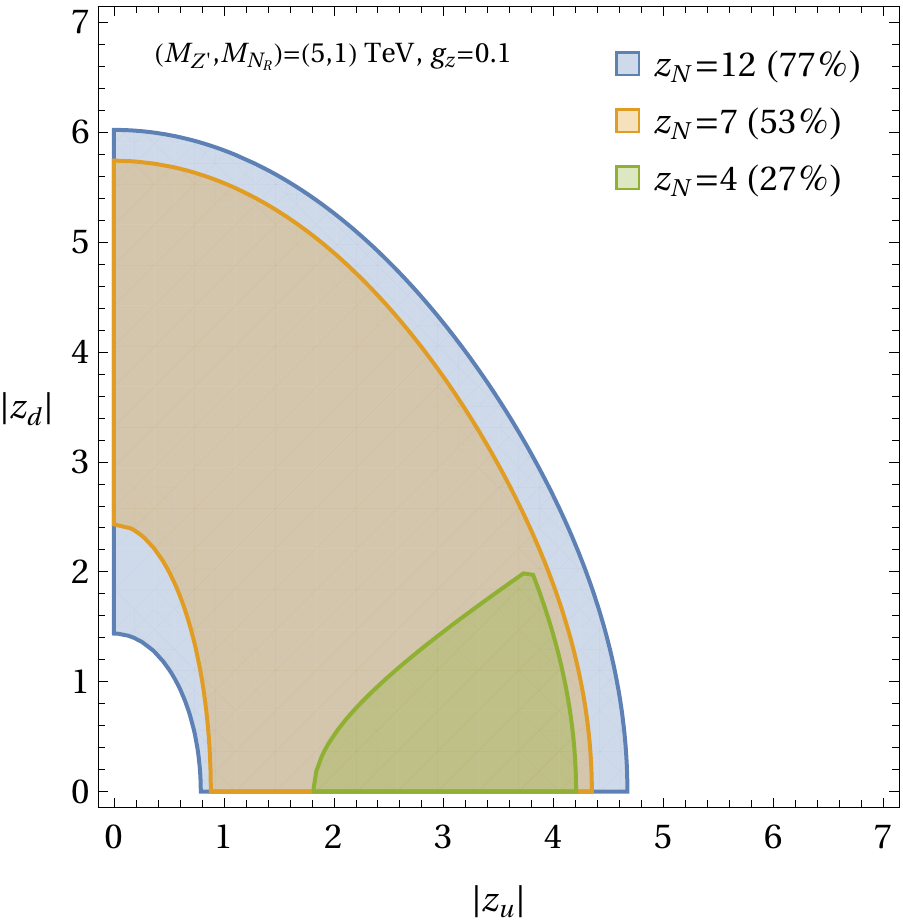}\label{fig:zuzdznHLLHC}}\hspace{1cm}
\subfloat[\quad\quad\quad(b)]{\includegraphics[height=6.5cm,width=7.5cm]{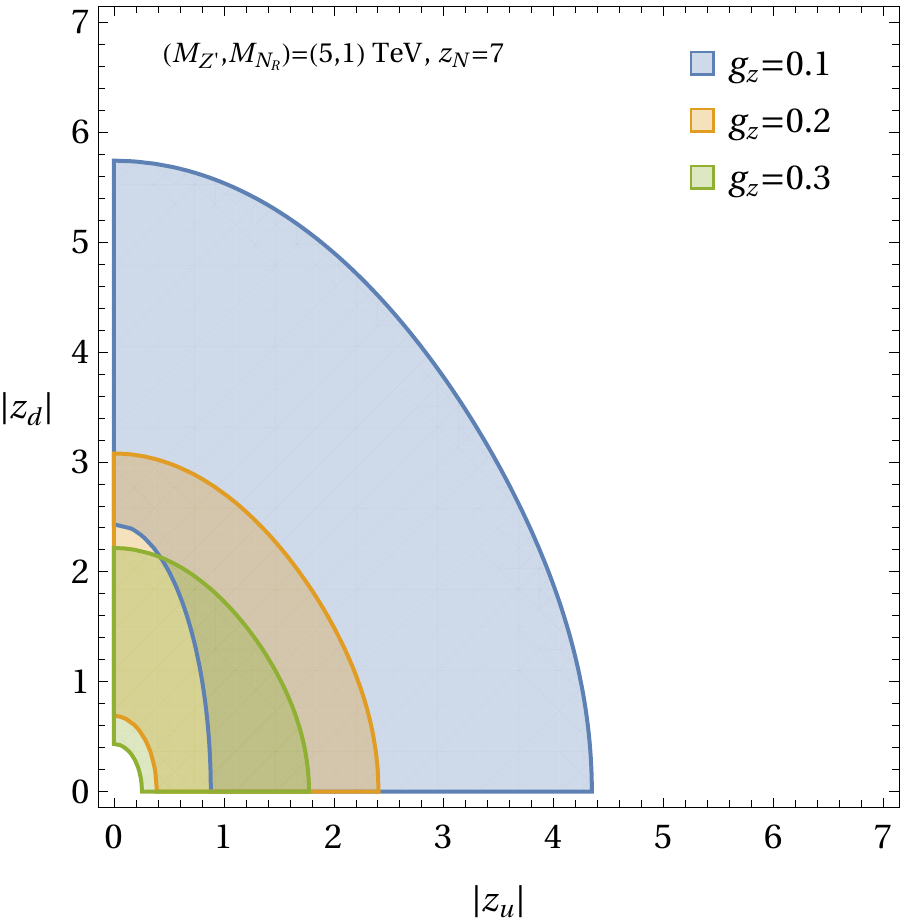}\label{fig:zuzdgxHLLHC}}\\
\subfloat[\quad\quad\quad(c)]{\includegraphics[height=6.5cm,width=7.5cm]{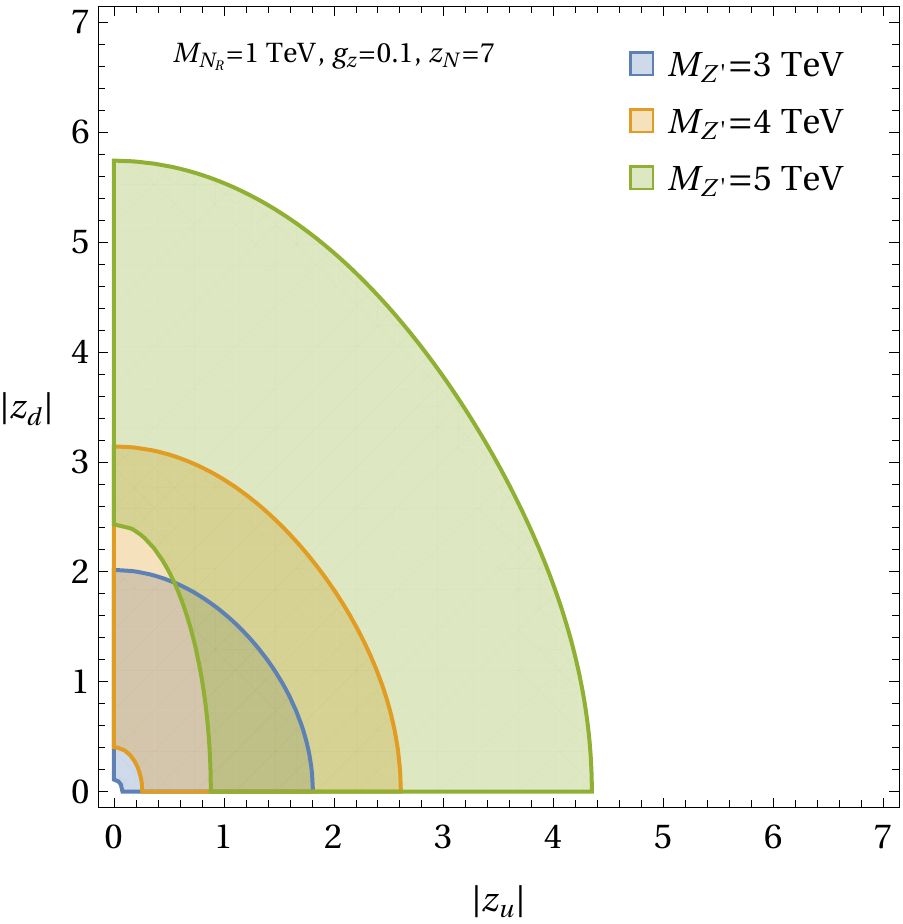}\label{fig:zuzdmzpHLLHC}}\hspace{1cm}
\subfloat[\quad\quad\quad(d)]{\includegraphics[height=6.5cm,width=7.5cm]{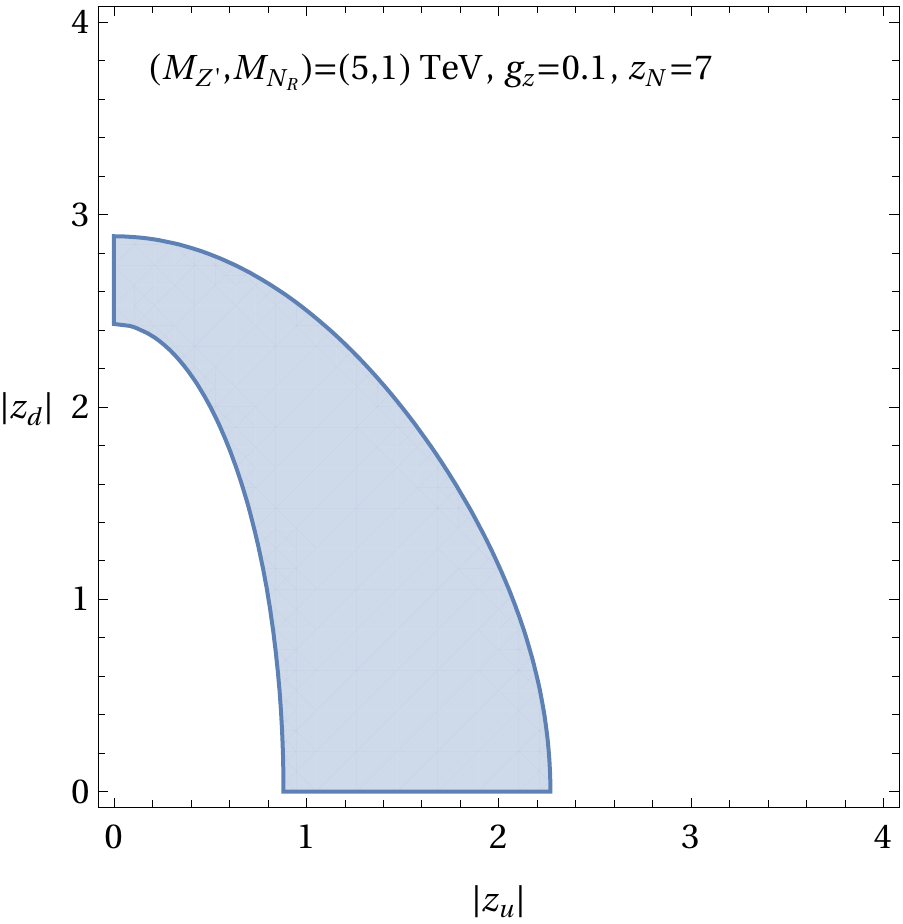}\label{fig:zuzdcomHLLHC}}
\caption{Regions in the $z_u-z_d$ plane that can be probed through the RHN-pair-production channel with more than $5\sg$ significance at the HL-LHC for 3000~$\textrm{fb}^{-1}$ integrated luminosity. These regions are allowed by the latest dijet search data as presented in Fig.~\ref{fig:zuzdalreg}. The region shown in (d) is beyond the projected reach of the dijet channel but can be probed with more than $5\sg$ significance using our channel.}
\label{fig:zuzdHLLHC}
\end{figure*} 
\begin{figure}
\captionsetup[subfigure]{labelformat=empty}
\subfloat[(a)]{\includegraphics[height=6.5cm,width=7.5cm]{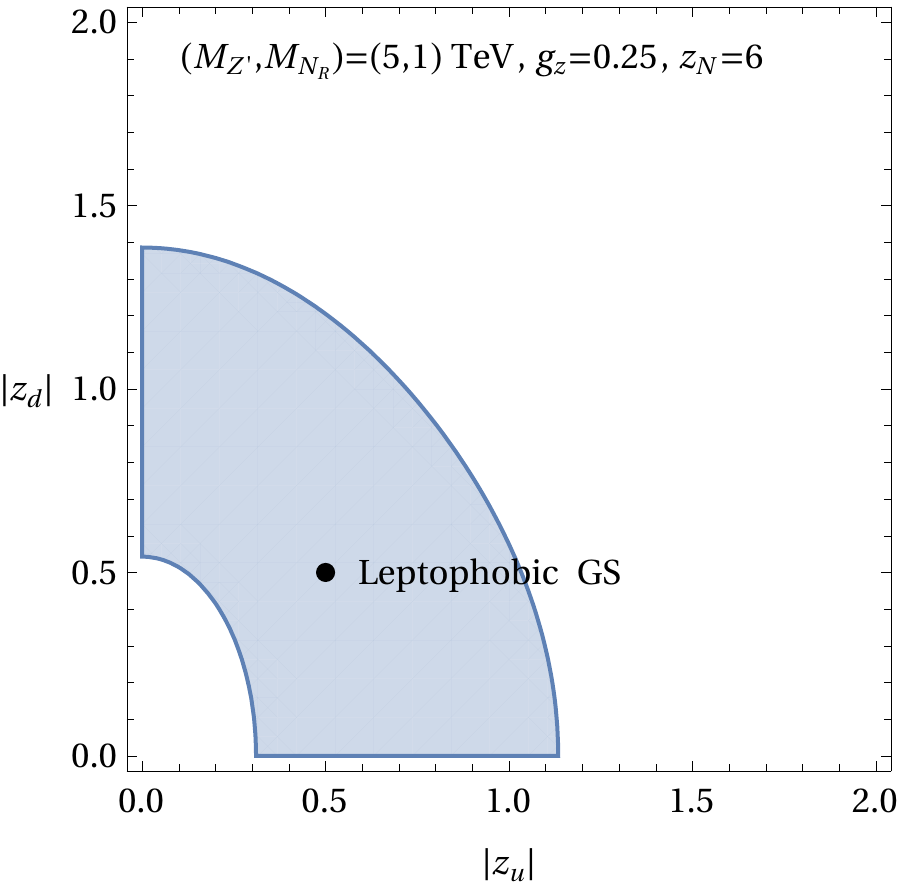}\label{fig:GSzuzd}}\\
\subfloat[(b)]{\includegraphics[height=6.5cm,width=7.5cm]{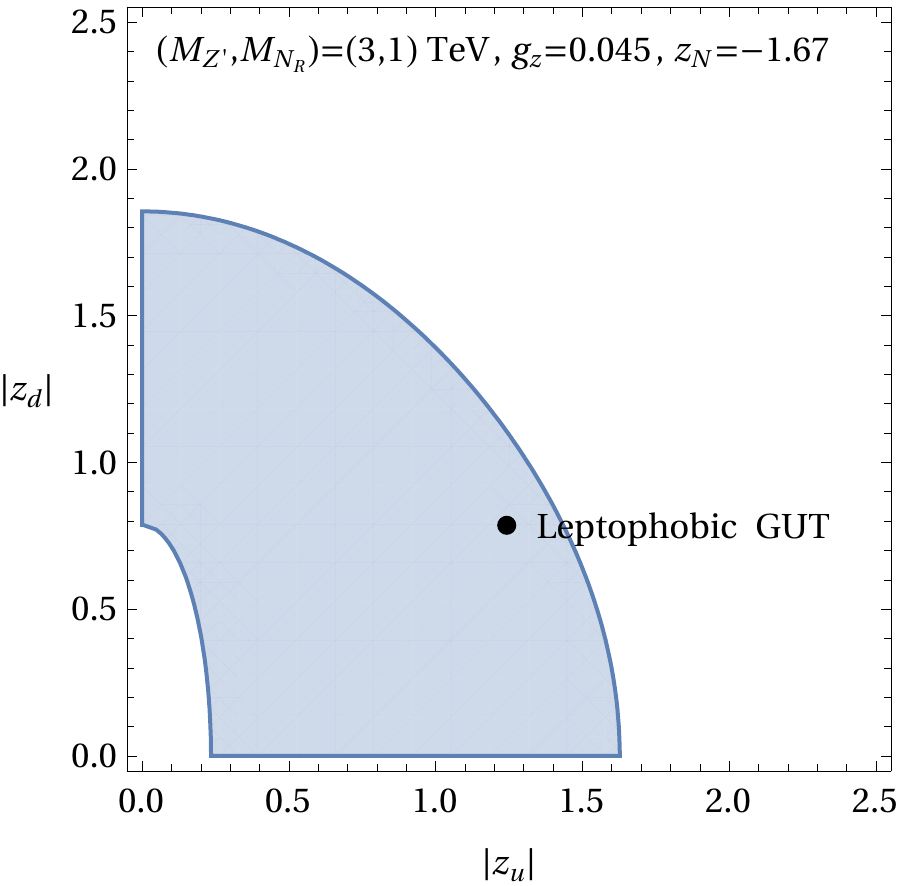}\label{fig:gutzuzd}}
\caption{Regions in $z_u-z_d$ plane that are beyond the projected HL-LHC reach in the dijet channel but can be probed with more than $5\sg$
significance using our channel. The back dots represent the (a) leptophobic GS and (b) leptophobic GUT models discussed in Sec.~\ref{sec:LPU1}.}
\label{fig:gsgutzuzd}
\end{figure}

\section{Prospects at the HL-LHC}
\label{sec:reach}
\noindent
Since, some of the background processes are large, we apply the following generation-level (preselection) cuts at the time of parton-level event generation to save computation time, 
\begin{equation*} 
p_{T}(\mu_{1}),p_{T}(\mu_{2}) > 100~\textrm{GeV};~M(\mu_1,\mu_2) > 120~\textrm{GeV}\ ,
\end{equation*}
where $p_{T}(\mu_1)$ and $p_T(\mu_2)$ are the transverse momenta of leading and subleading $p_T$-ordered muons and $M(\mu_1,\mu_2)$ is the invariant mass of the muon pair. These strong cuts affect both the signal and the background event generations but they drastically reduce the large backgrounds by about two-three orders of magnitude.
More advanced cuts are applied at the final selection level. To generate the $W+jets$ background, where two leptons are not present, we apply only the $p_T(\mu)>100$~GeV cut at the generation level.  

For our analysis, we apply the following selection cuts in sequence.
\begin{enumerate}
\item[$\mc C_1$.] Transverse momenta of the two muons should satisfy, $p_T(\mu_1) > 150~\textrm{GeV}$, $p_{T}(\mu_2) > 100~\textrm{GeV}$.

\item[$\mc C_2$.] Invariant mass of the muon pair, $M(\mu_1,\mu_2) > 200~\textrm{GeV}$.
     
\item[$\mc C_3$.] Fatjet mass, $|M(J_i) - m_W| < 40~\textrm{GeV}$, where $i$ goes up to $2$ in the events with two or more fatjets.

\item[$\mc C_4$.] In the events with only one fatjet, the invariant mass of one fatjet-muon pair should satisfy, $|M(J,\mu_i) - M_{N_{R}}| < 0.25\, M_{N_{R}}$ where $i=1$ or $2$. For the events with more fatjets, the leading two fatjets and muon pairs should satisfy the following two criteria, 
\begin{enumerate}
\item 
Both $|M(J_1,\mu_i) - M_{N_{R}}|$ and $|M(J_2,\mu_j) - M_{N_{R}}|$ must be less than $0.25\, M_{N_R}$, where $i \neq j$ and $i,j=1$ or $2$  and
\item
$M(J_1, J_2,\mu_1,\mu_2) > {\rm Min}(0.6\, M_{Z^\prime}, 1800~{\rm GeV})$.
\end{enumerate}
    
\item[$\mc C_5$.] For $M_{Z^\prime} \leq 3000$ GeV, the fatjet $N$-subjettiness ratio should satisfy, $\tau_{21}(J_i) < 0.6$ where $i$ goes up to $2$ when there are two or more fatjets present.

\item[$\mc C_6$.] The scalar sum of $p_T$ of all visible particles in the final state, $S_T > {\rm Min}(0.6\, M_{Z^\prime}, 1800~{\rm GeV})$.
\end{enumerate}

The effects of these cuts are shown in the cutflow table (Table~\ref{tab:cutflow}). There the numbers represent the number of events (at the $\sqrt{s}=14$~TeV LHC with $\mc{L}=3000~\textrm{fb}^{-1}$ of integrated luminosity) surviving after each cut. The table is generated for the benchmark parameters: $M_{Z^\prime}=3500$ GeV, $M_{N_R}=1000$ GeV, and $g_z=0.1$.

To estimate the contribution of jets faking as leptons to the background from the $W+jets$ process, we follow a simple method. We treat the leading or subleading AK4 jet as the second lepton. We ensure the fatjets do not overlap with the fake lepton. After applying all cuts and multiplying with a jet-faking efficiency of $10^{-4}$~\cite{Curtin:2013zua}, $W+jets$ contributes negligibly to the total background.

The signal significance is given by the $\mc{Z}$ score which can be estimated as
\begin{equation}
\label{Assimov}
\mathcal{Z} = \sqrt{2(N_S + N_B) \ln \left({{N_S + N_B}}\over{N_B} \right) - 2N_S}\ ,
\end{equation}
where $N_S$ ($N_B$) is the number of signal (background) events at the $\sqrt{s}=14$~TeV LHC surviving the set of cuts at $3000~\textrm{fb}^{-1}$ of integrated luminosity. Fig.~\ref{fig:signi} shows the $5\sg$ discovery and the $2\sg$ exclusion reaches in the $M_{Z^\prime}-M_{N_R}$ plane for a range of  the coupling $g_z$, defined in Eq.~\eqref{eq:modlag}.  The contours in Fig.~\ref{fig:signi} represent  particular values of $g_z$. There are three free charges present in our simplified model. In a strict sense, our results are valid only when the narrow-width approximation is valid. For a very large coupling, the large width would eventually affect the distributions and the discovery/exclusion reaches.

We present the regions that can be probed with the opposite-sign-dimuon channel with more than $5\sg$ significance at the HL-LHC in Fig.~\ref{fig:zuzdHLLHC}. These regions are allowed by the latest dijet resonance search data. It is interesting to note that large regions of the parameter space remain open for the channel considered here. When the luminosity increases in the future, the dijet bound would also improve. Since the current bounds are obtained for about $140$ fb$^{-1}$ of integrated luminosity, at the HL-LHC, the upper limit on the production cross section in the dijet resonance search channel will come down roughly by a factor $\sqrt{3000/140}\approx 4.65$. With this projected dijet bound, we see that large regions of the parameter space remain exclusively open for our signal channel. For example, in Fig.~\ref{fig:zuzdcomHLLHC}, we show a region in which the dijet channel cannot probe at the HL-LHC, but the RHN pair production through a $Z^\prime$ channel can. In our leptophobic GS and the leptophobic GUT models, the charges $z_{u,d,N}$ are fixed. In Fig.~\ref{fig:gsgutzuzd}, we show the regions in $z_u-z_d$ plane which can be probed exclusively at the HL-LHC with our signal for some benchmark choices of the free parameters (where $z_N$ values are motivated from the models discussed in Sec.~\ref{sec:LPU1}). 
In Figs.~\ref{fig:GSzuzd} and \ref{fig:gutzuzd}, we show the black dots representing the leptophobic GS and leptophopbic GUT points, respectively. 

\section{Summary and Conclusions}
\label{sec:conclu}
\noindent
We investigated the HL-LHC prospects of a leptophobic heavy neutral gauge boson $Z^\prime$ that decays to a RHN pair. The $Z^\prime \to N_RN_R$ channel has not been searched for in the LHC experiments. Since the RHNs are SM-gauge singlets, it is tough to produce them at the LHC. However, the production rates of the RHNs can be considerable if they are produced in the decay of another BSM particle, e.g., $W^\prime \to \ell N_R$ or $Z^\prime\to N_RN_R$, etc. Experimental searches of correlated $W^\prime$ and $N_R$ signatures have been performed before at the LHC, but not the correlated $Z^\prime$ and $N_R$ search; even though the $pp\to Z^\prime\to N_RN_R$ channel is present in many simple extensions of the SM, like the left-right models~\cite{Ferrari:2002ac}, or the anomaly-free $U(1)$ extensions~\cite{Das:2017flq,Das:2017deo}, etc. In these models, the $Z^\prime \to \ell^+\ell^-$ channel dominates almost in the entire parameter space, and hence, normally, one would not consider the $Z^\prime\to N_RN_R$ channel as a primary channel to probe the $Z^\prime$. However, the RHN channel becomes important in a leptophobic $Z^\prime$ model---it can even be the dominant channel if BR($Z^\prime\to N_RN_R$) is high. This is essentially the main novel aspect of our study.

One can realise a leptophobic $Z^\prime$ with high branching to RHN pairs using the GS anomaly-cancellation mechanism or within the GUT models. As a motivation, we illustrated two such examples where we get a leptophobic $Z^\prime$. We introduced a phenomenological Lagrangian representing a class of leptophobic $U(1)$ extensions. Our model-independent setup contains only a few free parameters relevant for the collider analysis (the new $U(1)$ gauge coupling, the three $U(1)$ charges, and the two masses). One can easily map the parameters onto a wide class of low-scale leptophobic $Z^\prime$ models containing the $Z^\prime\to N_RN_R$ decay mode and use our results directly. Of the final states arising from the subsequent decay of the RHNs in the  RHN-pair channel, we considered  the opposite-sign dilepton (dimuon) plus at least one $W$-like fatjet final state. The opposite-sign dilepton is a signature of the \emph{Dirac-type} RHNs or equivalently, the inverse-seesaw mechanism generating the neutrino masses. The same-flavour OSDL channel is vital for probing the RHNs produced either from a heavy gauge boson like the $Z^\prime$ or $W^\prime$~\cite{ThomasArun:2021rwf}. 

In the absence of the dilepton bounds, the major constraints on our model parameters come from the dijet resonance searches at the LHC. Considering the latest results from the LHC, we found that large regions of the parameter space are open for the $Z^\prime\to N_RN_R$ channel. We also found that this channel can probe regions beyond the reach of the dijet channel at the HL-LHC. 

\section*{Acknowledgments}
\noindent
We thank Cyrin Neeraj for drawing the Feynman diagram.  M.T.A. is financially supported by the DST through INSPIRE Faculty Grant No. DST/INSPIRE/$04$/$2019$/$002507$. The research of A.C. is supported by the European Regional Development Fund under Grant No. TK$133$. T.M. acknowledges the use of the high-performance computing facility at IISER-TVM.

\bibliography{zprime_reference}

\providecommand{\href}[2]{#2}\begingroup\raggedright\begin{thebibliography}{10}

\bibitem{Langacker:1980js}
P.~Langacker, \emph{{Grand Unified Theories and Proton Decay}},
  \href{http://dx.doi.org/10.1016/0370-1573(81)90059-4}{\emph{Phys. Rept.} {\bf
  72} (1981) 185}.

\bibitem{London:1986dk}
D.~London and J.~L. Rosner, \emph{{Extra Gauge Bosons in E(6)}},
  \href{http://dx.doi.org/10.1103/PhysRevD.34.1530}{\emph{Phys. Rev. D} {\bf
  34} (1986) 1530}.

\bibitem{Hewett:1988xc}
J.~L. Hewett and T.~G. Rizzo, \emph{{Low-Energy Phenomenology of Superstring
  Inspired E(6) Models}},
  \href{http://dx.doi.org/10.1016/0370-1573(89)90071-9}{\emph{Phys. Rept.} {\bf
  183} (1989) 193}.

\bibitem{Leike:1998wr}
A.~Leike, \emph{{The Phenomenology of extra neutral gauge bosons}},
  \href{http://dx.doi.org/10.1016/S0370-1573(98)00133-1}{\emph{Phys. Rept.}
  {\bf 317} (1999) 143--250}, [\href{http://arxiv.org/abs/hep-ph/9805494}{{\tt
  hep-ph/9805494}}].

\bibitem{Langacker:2008yv}
P.~Langacker, \emph{{The Physics of Heavy $Z^\prime$ Gauge Bosons}},
  \href{http://dx.doi.org/10.1103/RevModPhys.81.1199}{\emph{Rev. Mod. Phys.}
  {\bf 81} (2009) 1199--1228}, [\href{http://arxiv.org/abs/0801.1345}{{\tt
  0801.1345}}].

\bibitem{Ekstedt:2016wyi}
A.~Ekstedt, R.~Enberg, G.~Ingelman, J.~L\"ofgren and T.~Mandal,
  \emph{{Constraining minimal anomaly free $\mathrm{U}(1)$ extensions of the
  Standard Model}},
  \href{http://dx.doi.org/10.1007/JHEP11(2016)071}{\emph{JHEP} {\bf 11} (2016)
  071}, [\href{http://arxiv.org/abs/1605.04855}{{\tt 1605.04855}}].

\bibitem{Leontaris:1999wf}
G.~K. Leontaris and J.~Rizos, \emph{{New fermion mass textures from anomalous
  U(1) symmetries with baryon and lepton number conservation}},
  \href{http://dx.doi.org/10.1016/S0550-3213(99)00723-3}{\emph{Nucl. Phys. B}
  {\bf 567} (2000) 32--60}, [\href{http://arxiv.org/abs/hep-ph/9909206}{{\tt
  hep-ph/9909206}}].

\bibitem{Ekstedt:2017tbo}
A.~Ekstedt, R.~Enberg, G.~Ingelman, J.~L\"ofgren and T.~Mandal, \emph{{Minimal
  anomalous $\mathrm{U}(1)$ theories and collider phenomenology}},
  \href{http://dx.doi.org/10.1007/JHEP02(2018)152}{\emph{JHEP} {\bf 02} (2018)
  152}, [\href{http://arxiv.org/abs/1712.03410}{{\tt 1712.03410}}].

\bibitem{ATLAS:2019fgd}
{\bf ATLAS} collaboration, G.~Aad et~al., \emph{{Search for new resonances in
  mass distributions of jet pairs using 139 fb$^{-1}$ of $pp$ collisions at
  $\sqrt{s}=13$ TeV with the ATLAS detector}},
  \href{http://dx.doi.org/10.1007/JHEP03(2020)145}{\emph{JHEP} {\bf 03} (2020)
  145}, [\href{http://arxiv.org/abs/1910.08447}{{\tt 1910.08447}}].

\bibitem{CMS:2019gwf}
{\bf CMS} collaboration, A.~M. Sirunyan et~al., \emph{{Search for high mass
  dijet resonances with a new background prediction method in proton-proton
  collisions at $\sqrt{s} =$ 13 TeV}},
  \href{http://dx.doi.org/10.1007/JHEP05(2020)033}{\emph{JHEP} {\bf 05} (2020)
  033}, [\href{http://arxiv.org/abs/1911.03947}{{\tt 1911.03947}}].

\bibitem{ATLAS:2019erb}
{\bf ATLAS} collaboration, G.~Aad et~al., \emph{{Search for high-mass dilepton
  resonances using 139 fb$^{-1}$ of $pp$ collision data collected at
  $\sqrt{s}=$13 TeV with the ATLAS detector}},
  \href{http://dx.doi.org/10.1016/j.physletb.2019.07.016}{\emph{Phys. Lett. B}
  {\bf 796} (2019) 68--87}, [\href{http://arxiv.org/abs/1903.06248}{{\tt
  1903.06248}}].

\bibitem{CMS:2021ctt}
{\bf CMS} collaboration, A.~M. Sirunyan et~al., \emph{{Search for resonant and
  nonresonant new phenomena in high-mass dilepton final states at $ \sqrt{s} $
  = 13 TeV}}, \href{http://dx.doi.org/10.1007/JHEP07(2021)208}{\emph{JHEP} {\bf
  07} (2021) 208}, [\href{http://arxiv.org/abs/2103.02708}{{\tt 2103.02708}}].

\bibitem{ATLAS:2016gzy}
{\bf ATLAS} collaboration, M.~Aaboud et~al., \emph{{Search for resonances in
  diphoton events at $\sqrt{s}$=13 TeV with the ATLAS detector}},
  \href{http://dx.doi.org/10.1007/JHEP09(2016)001}{\emph{JHEP} {\bf 09} (2016)
  001}, [\href{http://arxiv.org/abs/1606.03833}{{\tt 1606.03833}}].

\bibitem{CMS:2016kgr}
{\bf CMS} collaboration, V.~Khachatryan et~al., \emph{{Search for high-mass
  diphoton resonances in proton\textendash{}proton collisions at 13 TeV and
  combination with 8 TeV search}},
  \href{http://dx.doi.org/10.1016/j.physletb.2017.01.027}{\emph{Phys. Lett. B}
  {\bf 767} (2017) 147--170}, [\href{http://arxiv.org/abs/1609.02507}{{\tt
  1609.02507}}].

\bibitem{ATLAS:2020fry}
{\bf ATLAS} collaboration, G.~Aad et~al., \emph{{Search for heavy diboson
  resonances in semileptonic final states in pp collisions at $\sqrt{s}=13$ TeV
  with the ATLAS detector}},
  \href{http://dx.doi.org/10.1140/epjc/s10052-020-08554-y}{\emph{Eur. Phys. J.
  C} {\bf 80} (2020) 1165}, [\href{http://arxiv.org/abs/2004.14636}{{\tt
  2004.14636}}].

\bibitem{CMS:2021klu}
{\bf CMS} collaboration, A.~Tumasyan et~al., \emph{{Search for heavy resonances
  decaying to WW, WZ, or WH boson pairs in the lepton plus merged jet final
  state in proton-proton collisions at $\sqrt{s}$ = 13 TeV}},
  \href{http://arxiv.org/abs/2109.06055}{{\tt 2109.06055}}.

\bibitem{ATLAS:2012dgv}
{\bf ATLAS} collaboration, G.~Aad et~al., \emph{{Search for resonances decaying
  into top-quark pairs using fully hadronic decays in $pp$ collisions with
  ATLAS at $\sqrt{s}=7$ TeV}},
  \href{http://dx.doi.org/10.1007/JHEP01(2013)116}{\emph{JHEP} {\bf 01} (2013)
  116}, [\href{http://arxiv.org/abs/1211.2202}{{\tt 1211.2202}}].

\bibitem{CMS:2015fhb}
{\bf CMS} collaboration, V.~Khachatryan et~al., \emph{{Search for resonant $t
  \bar t$ production in proton-proton collisions at $\sqrt s=$ 8 TeV}},
  \href{http://dx.doi.org/10.1103/PhysRevD.93.012001}{\emph{Phys. Rev. D} {\bf
  93} (2016) 012001}, [\href{http://arxiv.org/abs/1506.03062}{{\tt
  1506.03062}}].

\bibitem{ATLAS:2018tvr}
{\bf ATLAS} collaboration, \emph{{Prospects for searches for heavy $Z^\prime$
  and $W^\prime$ bosons in fermionic final states with the ATLAS experiment at
  the HL-LHC}}, .

\bibitem{Ferrari:2002ac}
A.~Ferrari, \emph{{Study of the production of a new $Z^\prime$ boson and its
  decay into Majorana neutrinos in $p p$ collisions at $s$ = 14-TeV and in
  $e^{+} e^{-}$ collisions at $s$ = 3-TeV}},
  \href{http://dx.doi.org/10.1103/PhysRevD.65.093008}{\emph{Phys. Rev. D} {\bf
  65} (2002) 093008}.

\bibitem{Das:2017flq}
A.~Das, N.~Okada and D.~Raut, \emph{{Enhanced pair production of heavy Majorana
  neutrinos at the LHC}},
  \href{http://dx.doi.org/10.1103/PhysRevD.97.115023}{\emph{Phys. Rev. D} {\bf
  97} (2018) 115023}, [\href{http://arxiv.org/abs/1710.03377}{{\tt
  1710.03377}}].

\bibitem{Das:2017deo}
A.~Das, N.~Okada and D.~Raut, \emph{{Heavy Majorana neutrino pair productions
  at the LHC in minimal U(1) extended Standard Model}},
  \href{http://dx.doi.org/10.1140/epjc/s10052-018-6171-8}{\emph{Eur. Phys. J.
  C} {\bf 78} (2018) 696}, [\href{http://arxiv.org/abs/1711.09896}{{\tt
  1711.09896}}].

\bibitem{Cox:2017eme}
P.~Cox, C.~Han and T.~T. Yanagida, \emph{{LHC Search for Right-handed Neutrinos
  in $Z^\prime$ Models}},
  \href{http://dx.doi.org/10.1007/JHEP01(2018)037}{\emph{JHEP} {\bf 01} (2018)
  037}, [\href{http://arxiv.org/abs/1707.04532}{{\tt 1707.04532}}].

\bibitem{Das:2022rbl}
A.~Das, S.~Mandal, T.~Nomura and S.~Shil, \emph{{Heavy Majorana neutrino pair
  production from $Z^\prime$ at hadron and lepton colliders}},
  \href{http://arxiv.org/abs/2202.13358}{{\tt 2202.13358}}.

\bibitem{Babu:1996vt}
K.~S. Babu, C.~F. Kolda and J.~March-Russell, \emph{{Leptophobic U(1) $s$ and
  the R($b$) - R($c$) crisis}},
  \href{http://dx.doi.org/10.1103/PhysRevD.54.4635}{\emph{Phys. Rev. D} {\bf
  54} (1996) 4635--4647}, [\href{http://arxiv.org/abs/hep-ph/9603212}{{\tt
  hep-ph/9603212}}].

\bibitem{Lopez:1996ta}
J.~L. Lopez and D.~V. Nanopoulos, \emph{{Leptophobic $Z^\prime$ in stringy
  flipped SU(5)}}, \href{http://dx.doi.org/10.1103/PhysRevD.55.397}{\emph{Phys.
  Rev. D} {\bf 55} (1997) 397--406},
  [\href{http://arxiv.org/abs/hep-ph/9605359}{{\tt hep-ph/9605359}}].

\bibitem{Rizzo:1998ut}
T.~G. Rizzo, \emph{{Gauge kinetic mixing and leptophobic $Z^\prime$ in E(6) and
  SO(10)}}, \href{http://dx.doi.org/10.1103/PhysRevD.59.015020}{\emph{Phys.
  Rev. D} {\bf 59} (1998) 015020},
  [\href{http://arxiv.org/abs/hep-ph/9806397}{{\tt hep-ph/9806397}}].

\bibitem{Leroux:2001fx}
K.~Leroux and D.~London, \emph{{Flavor changing neutral currents and
  leptophobic $Z^\prime$ gauge bosons}},
  \href{http://dx.doi.org/10.1016/S0370-2693(01)01489-7}{\emph{Phys. Lett. B}
  {\bf 526} (2002) 97--103}, [\href{http://arxiv.org/abs/hep-ph/0111246}{{\tt
  hep-ph/0111246}}].

\bibitem{Minkowski:1977sc}
P.~Minkowski, \emph{{$\mu \to e\gamma$ at a Rate of One Out of $10^{9}$ Muon
  Decays?}}, \href{http://dx.doi.org/10.1016/0370-2693(77)90435-X}{\emph{Phys.
  Lett. B} {\bf 67} (1977) 421--428}.

\bibitem{Mohapatra:1979ia}
R.~N. Mohapatra and G.~Senjanovic, \emph{{Neutrino Mass and Spontaneous Parity
  Nonconservation}},
  \href{http://dx.doi.org/10.1103/PhysRevLett.44.912}{\emph{Phys. Rev. Lett.}
  {\bf 44} (1980) 912}.

\bibitem{Mohapatra:1986aw}
R.~N. Mohapatra, \emph{{Mechanism for Understanding Small Neutrino Mass in
  Superstring Theories}},
  \href{http://dx.doi.org/10.1103/PhysRevLett.56.561}{\emph{Phys. Rev. Lett.}
  {\bf 56} (1986) 561--563}.

\bibitem{Mohapatra:1986bd}
R.~N. Mohapatra and J.~W.~F. Valle, \emph{{Neutrino Mass and Baryon Number
  Nonconservation in Superstring Models}},
  \href{http://dx.doi.org/10.1103/PhysRevD.34.1642}{\emph{Phys. Rev. D} {\bf
  34} (1986) 1642}.

\bibitem{Das:2017kkm}
D.~Das, K.~Ghosh, M.~Mitra and S.~Mondal, \emph{{Probing sterile neutrinos in
  the framework of inverse seesaw mechanism through leptoquark productions}},
  \href{http://dx.doi.org/10.1103/PhysRevD.97.015024}{\emph{Phys. Rev. D} {\bf
  97} (2018) 015024}, [\href{http://arxiv.org/abs/1708.06206}{{\tt
  1708.06206}}].

\bibitem{Jana:2019mez}
S.~Jana, P.~K. Vishnu and S.~Saad, \emph{{Minimal dirac neutrino mass models
  from $\hbox {U}(1)_{\mathrm{R}}$ gauge symmetry and left\textendash{}right
  asymmetry at colliders}},
  \href{http://dx.doi.org/10.1140/epjc/s10052-019-7441-9}{\emph{Eur. Phys. J.
  C} {\bf 79} (2019) 916}, [\href{http://arxiv.org/abs/1904.07407}{{\tt
  1904.07407}}].

\bibitem{Das:2019pua}
A.~Das, S.~Goswami, K.~N. Vishnudath and T.~Nomura, \emph{{Constraining a
  general U(1)$^\prime$ inverse seesaw model from vacuum stability, dark matter
  and collider}},
  \href{http://dx.doi.org/10.1103/PhysRevD.101.055026}{\emph{Phys. Rev. D} {\bf
  101} (2020) 055026}, [\href{http://arxiv.org/abs/1905.00201}{{\tt
  1905.00201}}].

\bibitem{Choudhury:2020cpm}
D.~Choudhury, K.~Deka, T.~Mandal and S.~Sadhukhan, \emph{{Neutrino and $Z'$
  phenomenology in an anomaly-free $\mathbf{U}(1)$ extension: role of
  higher-dimensional operators}},
  \href{http://dx.doi.org/10.1007/JHEP06(2020)111}{\emph{JHEP} {\bf 06} (2020)
  111}, [\href{http://arxiv.org/abs/2002.02349}{{\tt 2002.02349}}].

\bibitem{Bandyopadhyay:2020djh}
P.~Bandyopadhyay, S.~Jangid and M.~Mitra, \emph{{Scrutinizing Vacuum Stability
  in IDM with Type-III Inverse seesaw}},
  \href{http://dx.doi.org/10.1007/JHEP02(2021)075}{\emph{JHEP} {\bf 02} (2021)
  075}, [\href{http://arxiv.org/abs/2008.11956}{{\tt 2008.11956}}].

\bibitem{Deka:2021koh}
K.~Deka, T.~Mandal, A.~Mukherjee and S.~Sadhukhan, \emph{{Leptogenesis in an
  anomaly-free $\mathrm{U}(1)$ extension with higher-dimensional operators}},
  \href{http://arxiv.org/abs/2105.15088}{{\tt 2105.15088}}.

\bibitem{Das:2021nqj}
A.~Das, S.~Goswami, V.~K.~N. and T.~K. Poddar, \emph{{Freeze-in sterile
  neutrino dark matter in a class of U$(1)^\prime$ models with inverse
  seesaw}},  \href{http://arxiv.org/abs/2104.13986}{{\tt 2104.13986}}.

\bibitem{Banerjee:2015gca}
S.~Banerjee, P.~S.~B. Dev, A.~Ibarra, T.~Mandal and M.~Mitra, \emph{{Prospects
  of Heavy Neutrino Searches at Future Lepton Colliders}},
  \href{http://dx.doi.org/10.1103/PhysRevD.92.075002}{\emph{Phys. Rev. D} {\bf
  92} (2015) 075002}, [\href{http://arxiv.org/abs/1503.05491}{{\tt
  1503.05491}}].

\bibitem{Chakraborty:2018khw}
S.~Chakraborty, M.~Mitra and S.~Shil, \emph{{Fat Jet Signature of a Heavy
  Neutrino at Lepton Collider}},
  \href{http://dx.doi.org/10.1103/PhysRevD.100.015012}{\emph{Phys. Rev. D} {\bf
  100} (2019) 015012}, [\href{http://arxiv.org/abs/1810.08970}{{\tt
  1810.08970}}].

\bibitem{Barducci:2022hll}
D.~Barducci and E.~Bertuzzo, \emph{{The see-saw portal at future Higgs
  factories: the role of dimension six operators}},
  \href{http://arxiv.org/abs/2201.11754}{{\tt 2201.11754}}.

\bibitem{Han:2006ip}
T.~Han and B.~Zhang, \emph{{Signatures for Majorana neutrinos at hadron
  colliders}},
  \href{http://dx.doi.org/10.1103/PhysRevLett.97.171804}{\emph{Phys. Rev.
  Lett.} {\bf 97} (2006) 171804},
  [\href{http://arxiv.org/abs/hep-ph/0604064}{{\tt hep-ph/0604064}}].

\bibitem{Chen:2011hc}
C.-Y. Chen and P.~S.~B. Dev, \emph{{Multi-Lepton Collider Signatures of Heavy
  Dirac and Majorana Neutrinos}},
  \href{http://dx.doi.org/10.1103/PhysRevD.85.093018}{\emph{Phys. Rev. D} {\bf
  85} (2012) 093018}, [\href{http://arxiv.org/abs/1112.6419}{{\tt 1112.6419}}].

\bibitem{Das:2017hmg}
A.~Das, P.~S.~B. Dev and R.~N. Mohapatra, \emph{{Same Sign versus Opposite Sign
  Dileptons as a Probe of Low Scale Seesaw Mechanisms}},
  \href{http://dx.doi.org/10.1103/PhysRevD.97.015018}{\emph{Phys. Rev. D} {\bf
  97} (2018) 015018}, [\href{http://arxiv.org/abs/1709.06553}{{\tt
  1709.06553}}].

\bibitem{Anastasopoulos:2006cz}
P.~Anastasopoulos, M.~Bianchi, E.~Dudas and E.~Kiritsis, \emph{{Anomalies,
  anomalous U(1)'s and generalized Chern-Simons terms}},
  \href{http://dx.doi.org/10.1088/1126-6708/2006/11/057}{\emph{JHEP} {\bf 11}
  (2006) 057}, [\href{http://arxiv.org/abs/hep-th/0605225}{{\tt
  hep-th/0605225}}].

\bibitem{Anastasopoulos:2008jt}
P.~Anastasopoulos, F.~Fucito, A.~Lionetto, G.~Pradisi, A.~Racioppi and Y.~S.
  Stanev, \emph{{Minimal Anomalous U(1)-prime Extension of the MSSM}},
  \href{http://dx.doi.org/10.1103/PhysRevD.78.085014}{\emph{Phys. Rev. D} {\bf
  78} (2008) 085014}, [\href{http://arxiv.org/abs/0804.1156}{{\tt 0804.1156}}].

\bibitem{Bhardwaj:2018lma}
A.~Bhardwaj, A.~Das, P.~Konar and A.~Thalapillil, \emph{{Looking for Minimal
  Inverse Seesaw scenarios at the LHC with Jet Substructure Techniques}},
  \href{http://dx.doi.org/10.1088/1361-6471/ab7769}{\emph{J. Phys. G} {\bf 47}
  (2020) 075002}, [\href{http://arxiv.org/abs/1801.00797}{{\tt 1801.00797}}].

\bibitem{Das:2017pvt}
A.~Das, \emph{{Pair production of heavy neutrinos in next-to-leading order QCD
  at the hadron colliders in the inverse seesaw framework}},
  \href{http://dx.doi.org/10.1142/S0217751X21500123}{\emph{Int. J. Mod. Phys.
  A} {\bf 36} (2021) 2150012}, [\href{http://arxiv.org/abs/1701.04946}{{\tt
  1701.04946}}].

\bibitem{Das:2012ze}
A.~Das and N.~Okada, \emph{{Inverse seesaw neutrino signatures at the LHC and
  ILC}}, \href{http://dx.doi.org/10.1103/PhysRevD.88.113001}{\emph{Phys. Rev.
  D} {\bf 88} (2013) 113001}, [\href{http://arxiv.org/abs/1207.3734}{{\tt
  1207.3734}}].

\bibitem{PhysRevD.99.123508}
M.~J. Dolan, T.~P. Dutka and R.~R. Volkas, \emph{Low-scale leptogenesis with
  minimal lepton flavor violation},
  \href{http://dx.doi.org/10.1103/PhysRevD.99.123508}{\emph{Phys. Rev. D} {\bf
  99} (Jun, 2019) 123508}.

\bibitem{Dolan:2018qpy}
M.~J. Dolan, T.~P. Dutka and R.~R. Volkas, \emph{{Dirac-Phase Thermal
  Leptogenesis in the extended Type-I Seesaw Model}},
  \href{http://dx.doi.org/10.1088/1475-7516/2018/06/012}{\emph{JCAP} {\bf 06}
  (2018) 012}, [\href{http://arxiv.org/abs/1802.08373}{{\tt 1802.08373}}].

\bibitem{Planck:2018vyg}
{\bf Planck} collaboration, N.~Aghanim et~al., \emph{{Planck 2018 results. VI.
  Cosmological parameters}},
  \href{http://dx.doi.org/10.1051/0004-6361/201833910}{\emph{Astron.
  Astrophys.} {\bf 641} (2020) A6},
  [\href{http://arxiv.org/abs/1807.06209}{{\tt 1807.06209}}]. [Erratum:
  Astron.Astrophys. 652, C4 (2021)].

\bibitem{pdg_2020}
P.~D. Group and Z.~{\it et al}, \emph{{Review of Particle Physics}},
  \href{http://dx.doi.org/10.1093/ptep/ptaa104}{\emph{Progress of Theoretical
  and Experimental Physics} {\bf 2020} (08, 2020) }. 083C01.

\bibitem{Casas:2001sr}
J.~A. Casas and A.~Ibarra, \emph{{Oscillating neutrinos and $\mu \to e,
  \gamma$}}, \href{http://dx.doi.org/10.1016/S0550-3213(01)00475-8}{\emph{Nucl.
  Phys. B} {\bf 618} (2001) 171--204},
  [\href{http://arxiv.org/abs/hep-ph/0103065}{{\tt hep-ph/0103065}}].

\bibitem{Mukherjee:2022fjm}
A.~Mukherjee and N.~Narendra, \emph{{Retrieving Inverse Seesaw parameter space
  for Dirac Phase Leptogenesis}},  \href{http://arxiv.org/abs/2204.08820}{{\tt
  2204.08820}}.

\bibitem{Eichten:1984eu}
E.~Eichten, I.~Hinchliffe, K.~D. Lane and C.~Quigg, \emph{{Super Collider
  Physics}}, \href{http://dx.doi.org/10.1103/RevModPhys.56.579}{\emph{Rev. Mod.
  Phys.} {\bf 56} (1984) 579--707}. [Addendum: Rev.Mod.Phys. 58, 1065--1073
  (1986)].

\bibitem{Kang:2015uoc}
Z.~Kang, P.~Ko and J.~Li, \emph{{New Avenues to Heavy Right-handed Neutrinos
  with Pair Production at Hadronic Colliders}},
  \href{http://dx.doi.org/10.1103/PhysRevD.93.075037}{\emph{Phys. Rev. D} {\bf
  93} (2016) 075037}, [\href{http://arxiv.org/abs/1512.08373}{{\tt
  1512.08373}}].

\bibitem{Accomando:2017qcs}
E.~Accomando, L.~Delle~Rose, S.~Moretti, E.~Olaiya and C.~H.
  Shepherd-Themistocleous, \emph{{Extra Higgs boson and $Z'$ as portals to
  signatures of heavy neutrinos at the LHC}},
  \href{http://dx.doi.org/10.1007/JHEP02(2018)109}{\emph{JHEP} {\bf 02} (2018)
  109}, [\href{http://arxiv.org/abs/1708.03650}{{\tt 1708.03650}}].

\bibitem{Huitu:2008gf}
K.~Huitu, S.~Khalil, H.~Okada and S.~K. Rai, \emph{{Signatures for right-handed
  neutrinos at the Large Hadron Collider}},
  \href{http://dx.doi.org/10.1103/PhysRevLett.101.181802}{\emph{Phys. Rev.
  Lett.} {\bf 101} (2008) 181802}, [\href{http://arxiv.org/abs/0803.2799}{{\tt
  0803.2799}}].

\bibitem{Deppisch:2019kvs}
F.~Deppisch, S.~Kulkarni and W.~Liu, \emph{{Heavy neutrino production via $Z'$
  at the lifetime frontier}},
  \href{http://dx.doi.org/10.1103/PhysRevD.100.035005}{\emph{Phys. Rev. D} {\bf
  100} (2019) 035005}, [\href{http://arxiv.org/abs/1905.11889}{{\tt
  1905.11889}}].

\bibitem{Das:2019fee}
A.~Das, P.~S.~B. Dev and N.~Okada, \emph{{Long-lived TeV-scale right-handed
  neutrino production at the LHC in gauged $U(1)_X$ model}},
  \href{http://dx.doi.org/10.1016/j.physletb.2019.135052}{\emph{Phys. Lett. B}
  {\bf 799} (2019) 135052}, [\href{http://arxiv.org/abs/1906.04132}{{\tt
  1906.04132}}].

\bibitem{Chiang:2019ajm}
C.-W. Chiang, G.~Cottin, A.~Das and S.~Mandal, \emph{{Displaced heavy neutrinos
  from $Z'$ decays at the LHC}},
  \href{http://dx.doi.org/10.1007/JHEP12(2019)070}{\emph{JHEP} {\bf 12} (2019)
  070}, [\href{http://arxiv.org/abs/1908.09838}{{\tt 1908.09838}}].

\bibitem{Padhan:2022fak}
R.~Padhan, M.~Mitra, S.~Kulkarni and F.~F. Deppisch, \emph{{Displaced fat-jets
  and tracks to probe boosted right-handed neutrinos in the $U(1)_{B-L}$
  model}},  3, 2022.
\newblock \href{http://arxiv.org/abs/2203.06114}{{\tt 2203.06114}}.

\bibitem{Degrande:2011ua}
C.~Degrande, C.~Duhr, B.~Fuks, D.~Grellscheid, O.~Mattelaer and T.~Reiter,
  \emph{{UFO - The Universal FeynRules Output}},
  \href{http://dx.doi.org/10.1016/j.cpc.2012.01.022}{\emph{Comput. Phys.
  Commun.} {\bf 183} (2012) 1201--1214},
  [\href{http://arxiv.org/abs/1108.2040}{{\tt 1108.2040}}].

\bibitem{Alloul:2013bka}
A.~Alloul, N.~D. Christensen, C.~Degrande, C.~Duhr and B.~Fuks,
  \emph{{FeynRules 2.0 - A complete toolbox for tree-level phenomenology}},
  \href{http://dx.doi.org/10.1016/j.cpc.2014.04.012}{\emph{Comput. Phys.
  Commun.} {\bf 185} (2014) 2250--2300},
  [\href{http://arxiv.org/abs/1310.1921}{{\tt 1310.1921}}].

\bibitem{Alwall:2014hca}
J.~Alwall, R.~Frederix, S.~Frixione, V.~Hirschi, F.~Maltoni, O.~Mattelaer
  et~al., \emph{{The automated computation of tree-level and next-to-leading
  order differential cross sections, and their matching to parton shower
  simulations}}, \href{http://dx.doi.org/10.1007/JHEP07(2014)079}{\emph{JHEP}
  {\bf 07} (2014) 079}, [\href{http://arxiv.org/abs/1405.0301}{{\tt
  1405.0301}}].

\bibitem{Ball:2012cx}
R.~D. Ball et~al., \emph{{Parton distributions with LHC data}},
  \href{http://dx.doi.org/10.1016/j.nuclphysb.2012.10.003}{\emph{Nucl. Phys. B}
  {\bf 867} (2013) 244--289}, [\href{http://arxiv.org/abs/1207.1303}{{\tt
  1207.1303}}].

\bibitem{Sjostrand:2014zea}
T.~Sj\"ostrand, S.~Ask, J.~R. Christiansen, R.~Corke, N.~Desai, P.~Ilten
  et~al., \emph{{An introduction to PYTHIA 8.2}},
  \href{http://dx.doi.org/10.1016/j.cpc.2015.01.024}{\emph{Comput. Phys.
  Commun.} {\bf 191} (2015) 159--177},
  [\href{http://arxiv.org/abs/1410.3012}{{\tt 1410.3012}}].

\bibitem{deFavereau:2013fsa}
{\bf DELPHES 3} collaboration, J.~de~Favereau, C.~Delaere, P.~Demin,
  A.~Giammanco, V.~Lema\^\i{}tre, A.~Mertens et~al., \emph{{DELPHES 3, A
  modular framework for fast simulation of a generic collider experiment}},
  \href{http://dx.doi.org/10.1007/JHEP02(2014)057}{\emph{JHEP} {\bf 02} (2014)
  057}, [\href{http://arxiv.org/abs/1307.6346}{{\tt 1307.6346}}].

\bibitem{Cacciari:2008gp}
M.~Cacciari, G.~P. Salam and G.~Soyez, \emph{{The anti-$k_t$ jet clustering
  algorithm}},
  \href{http://dx.doi.org/10.1088/1126-6708/2008/04/063}{\emph{JHEP} {\bf 04}
  (2008) 063}, [\href{http://arxiv.org/abs/0802.1189}{{\tt 0802.1189}}].

\bibitem{Cacciari:2011ma}
M.~Cacciari, G.~P. Salam and G.~Soyez, \emph{{FastJet User Manual}},
  \href{http://dx.doi.org/10.1140/epjc/s10052-012-1896-2}{\emph{Eur. Phys. J.
  C} {\bf 72} (2012) 1896}, [\href{http://arxiv.org/abs/1111.6097}{{\tt
  1111.6097}}].

\bibitem{Bhaskar:2021gsy}
A.~Bhaskar, T.~Mandal, S.~Mitra and M.~Sharma, \emph{{Improving
  third-generation leptoquark searches with combined signals and boosted top
  quarks}}, \href{http://dx.doi.org/10.1103/PhysRevD.104.075037}{\emph{Phys.
  Rev. D} {\bf 104} (2021) 075037},
  [\href{http://arxiv.org/abs/2106.07605}{{\tt 2106.07605}}].

\bibitem{Catani:2009sm}
S.~Catani, L.~Cieri, G.~Ferrera, D.~de~Florian and M.~Grazzini, \emph{{Vector
  boson production at hadron colliders: a fully exclusive QCD calculation at
  NNLO}}, \href{http://dx.doi.org/10.1103/PhysRevLett.103.082001}{\emph{Phys.
  Rev. Lett.} {\bf 103} (2009) 082001},
  [\href{http://arxiv.org/abs/0903.2120}{{\tt 0903.2120}}].

\bibitem{Balossini:2009sa}
G.~Balossini, G.~Montagna, C.~M. Carloni~Calame, M.~Moretti, O.~Nicrosini,
  F.~Piccinini et~al., \emph{{Combination of electroweak and QCD corrections to
  single W production at the Fermilab Tevatron and the CERN LHC}},
  \href{http://dx.doi.org/10.1007/JHEP01(2010)013}{\emph{JHEP} {\bf 01} (2010)
  013}, [\href{http://arxiv.org/abs/0907.0276}{{\tt 0907.0276}}].

\bibitem{Muselli:2015kba}
C.~Muselli, M.~Bonvini, S.~Forte, S.~Marzani and G.~Ridolfi, \emph{{Top Quark
  Pair Production beyond NNLO}},
  \href{http://dx.doi.org/10.1007/JHEP08(2015)076}{\emph{JHEP} {\bf 08} (2015)
  076}, [\href{http://arxiv.org/abs/1505.02006}{{\tt 1505.02006}}].

\bibitem{Kidonakis:2015nna}
N.~Kidonakis, \emph{{Theoretical results for electroweak-boson and single-top
  production}}, \href{http://dx.doi.org/10.22323/1.247.0170}{\emph{PoS} {\bf
  DIS2015} (2015) 170}, [\href{http://arxiv.org/abs/1506.04072}{{\tt
  1506.04072}}].

\bibitem{Campbell:2011bn}
J.~M. Campbell, R.~K. Ellis and C.~Williams, \emph{{Vector boson pair
  production at the LHC}},
  \href{http://dx.doi.org/10.1007/JHEP07(2011)018}{\emph{JHEP} {\bf 07} (2011)
  018}, [\href{http://arxiv.org/abs/1105.0020}{{\tt 1105.0020}}].

\bibitem{Kulesza:2018tqz}
A.~Kulesza, L.~Motyka, D.~Schwartl\"ander, T.~Stebel and V.~Theeuwes,
  \emph{{Associated production of a top quark pair with a heavy electroweak
  gauge boson at NLO$+$NNLL accuracy}},
  \href{http://dx.doi.org/10.1140/epjc/s10052-019-6746-z}{\emph{Eur. Phys. J.
  C} {\bf 79} (2019) 249}, [\href{http://arxiv.org/abs/1812.08622}{{\tt
  1812.08622}}].

\bibitem{Bhaskar:2020gkk}
A.~Bhaskar, T.~Mandal and S.~Mitra, \emph{{Boosting vector leptoquark searches
  with boosted tops}},
  \href{http://dx.doi.org/10.1103/PhysRevD.101.115015}{\emph{Phys. Rev. D} {\bf
  101} (2020) 115015}, [\href{http://arxiv.org/abs/2004.01096}{{\tt
  2004.01096}}].

\bibitem{Curtin:2013zua}
D.~Curtin, J.~Galloway and J.~G. Wacker, \emph{{Measuring the $t \bar th$
  coupling from same-sign dilepton $+2b$ measurements}},
  \href{http://dx.doi.org/10.1103/PhysRevD.88.093006}{\emph{Phys. Rev. D} {\bf
  88} (2013) 093006}, [\href{http://arxiv.org/abs/1306.5695}{{\tt 1306.5695}}].

\bibitem{ThomasArun:2021rwf}
M.~Thomas~Arun, T.~Mandal, S.~Mitra, A.~Mukherjee, L.~Priya and A.~Sampath,
  \emph{{Testing left-right symmetry with an inverse seesaw mechanism at the
  LHC}}, \href{http://dx.doi.org/10.1103/PhysRevD.105.115007}{\emph{Phys. Rev.
  D} {\bf 105} (2022) 115007}, [\href{http://arxiv.org/abs/2109.09585}{{\tt
  2109.09585}}].

\end{thebibliography}\endgroup
\bibliographystyle{JHEPCust}

\end{document}